\documentclass[journal=jacsat,manuscript=article]{achemso}

\usepackage{chemformula} 
\usepackage[T1]{fontenc} 
\usepackage[utf8x]{inputenc}

\usepackage{xcolor}

\author{J. W. Gonz\'alez}
\email{jhon.gonzalez@usm.cl}
\affiliation[UTFSM]
{Departamento de F\'{i}sica, Universidad 
	T\'{e}cnica Federico Santa Mar\'{i}a, Casilla Postal 
	110V, Valpara\'{i}so, Chile.}
	
\author{E. Fl\'orez}
\affiliation[UdeM]
{Facultad de Ciencias Básicas, Universidad de  Medellín,  Medellín, Colombia}

\author{J. D. Correa}
\email{jcorrea@udemedellin.edu.co}
\affiliation[UdeM]
{Facultad de Ciencias Básicas, Universidad de  Medellín,  Medellín, Colombia}

\title[Li-/Na-ion batteries: MoS$_2$]
  {MoS$_2$ 2D-polymorphs as Li-/Na-ion batteries: 1T' vs 2H phases} 

\abbreviations{IR,NMR,UV}

\begin{document}

\begin{tocentry}

	\includegraphics[clip,width=8.cm]{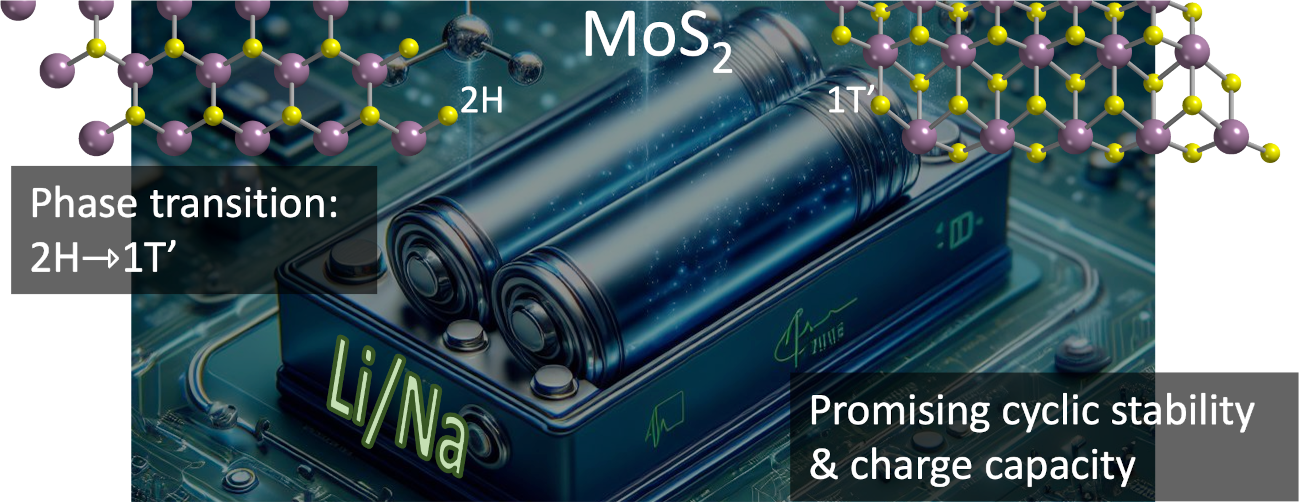}

%
%
%

\end{tocentry}

\begin{abstract}
In this study, we compare the performance of two phases of MoS$_2$ monolayers: 
1T' and 2H, about their ability to adsorb lithium and sodium ions. 
Employing the density functional theory and molecular dynamics, we include 
the ion concentration to analyze the electronic structure, ion kinetics, and battery performance.
The pristine 2H-MoS$_2$ monolayer is the ground state. However, the charge 
transfer effects above a critical ion concentration yields a stability change, 
where the 1T'-MoS$_2$ monolayer with adsorbed ions becomes more stable than 
the 2H counterpart. The diffusion of ions onto the 1T' monolayer is anisotropic, 
being more efficient at ion adsorption than the 2H phase. Finally, we calculate 
the open circuit voltage and specific capacity, confirming that the 1T'-MoS$_2$ 
phase has great potential for developing lithium/sodium ion batteries.   
\end{abstract}

\section{Introduction}
Recent advances in two-dimensional material research have led to the prediction and discovery of many new materials and phases with exciting phenomena\cite{dai2019strain,zhu2017recent,glavin2020emerging,gjerding2021recent}. Recently Transition metal dichalcogenides (TMDC)  have drawn great interest due to their large abundances, low costs, and unique structure, making them excellent candidates for studying fundamental physics phenomena and multiple applications in nanoelectronics\cite{manzeli20172d}. TMDC exists in several crystallographic phases\cite{sokolikova2020direct}, including 2H, 1T, 1T', 3R, and T$_d$\cite{wang2018strategies}. As an of the most striking of TMDs is molybdenum disulfide (MoS$_2$) due to its unique two-dimensional (2D) layer structure. In the case of MoS$_2$, a monolayer has three structural polymorphs. The 2H phase is the most typical, with trigonal-prismatic coordination of transition metal atoms and a semiconductor character. The 1T and 1T' phases, with octahedral coordination, are next in stability order and exhibit metallic character\cite{yu2018high}. However, the octahedral 1T phase is metastable and can decay spontaneously into a more stable distorted octahedral 1T'-phase\cite{sandoval1991raman,huang2018first}.

Two-dimensional materials are highly susceptible to changes in crystalline symmetry and can undergo phase transitions that lead to modifications in their physical and chemical properties\cite{lin2014atomic,huang2018first,yu2018high}. In monolayers of dichalcogenides, even weak external stimuli can cause polymorphic phase transitions. Previous studies show that charge transfer of metal atoms can stabilize the 1T' phase, reducing the energy barrier between the 2H and 1T' structures. This makes the 1T' phase the most stable configuration for adsorbed metal atoms at high saturation. To synthesize 1T'    -MoS$_2$, the phase transition engineering method is used experimentally. This method involves inducing a structural transition from the 2H to 1T phase by inserting lithium atoms with a concentration of 20\% and then obtaining 1T' phases when the Li atoms are deinserted\cite{hou2022phase}.

On the other hand, pursuing alternative materials capable of storing alkali ions aside from lithium is a thrilling and pressing matter, given that lithium is a scarce resource in high demand\cite{xu2020future,hajiahmadi2023first}. Sodium is an alternative to lithium due to its low cost, abundant resources, and similar storage mechanism to lithium\cite{sun2015vanadium}. \textcolor{black}{Due to the technological challenge of storing energy efficiently, some bases carbon materials have been studied as possible anode materials for  Li/Na batteries, showing theoretical capacities of 496.2 mAh/g for Na on twin-graphene\cite{dua2021twin} and 680 mAh/g for  Li/Na adsorbed on pentagraphyne (PG-yne)\cite{deb2022two}. Likewise, recent studies have found that two-dimensional TMDCs such as MoS$_2$ have better potential than graphite for alkali ion batteries due to their larger interlayer spacing, which can accommodate larger ions such as sodium with higher theoretical capacity\cite{lin2019two,upadhyay2021recent}.} 
The larger storage capacity of TMDCs led to an increased interest in studying the adsorption of different alkali atoms on several phases of MoS$_2$. It found that the adsorption of Li and Na on 2H and 1T' MoS$_2$ phases shows charge transfer. However, it has been observed that as the concentration of lithium ions increases, their ionicity decreases, and the ions slightly increase in ionic radius\cite{stephenson2014lithium}. In 2H-MoS$_2$ monolayers, inducing different types of vacancies enhances the system's performance for storing Li and Na ions\cite{barik2019defect}. Meanwhile, DFT calculations suggest that the 1T-MoS$_2$ monolayer is a promising anode material for Li, Na, and Mg ion batteries due to its high capacity, low open circuit voltage, and ultra-high ion diffusion kinetics\cite{he20221t}.  
However, recent research has shown that by incorporating Li ions into 2H-MoS$_2$, it can transform into 1T'-MoS$_2$, which greatly enhances its capacity to store Li ions, improves its cyclic stability, and increases its specific capacity in comparison to the 2H phases\cite{hou2022phase}. 

The progress of lithium-ion batteries (LIBs) has significantly enhanced rechargeable battery technology in numerous areas. Research has demonstrated that two-dimensional TMDCs such as MoS$_2$ may possess better potential than graphite for alkali ion batteries owing to their larger interlayer spacings, which can accommodate larger ions like sodium, and higher theoretical capacity. Our study delves into the feasibility of using the 1T'-MoS$_2$ phase as an anode material for Li and Na ions and comparing its performance with the 2H-MoS$_2$ phase, which has already demonstrated good performance for Li-ion batteries\cite{stephenson2014lithium}. We conducted ab-initio molecular dynamics to determine the structural and thermal stability of the monolayers, considering different Li/Na concentrations. Furthermore, we calculated the adsorption energy in gas/solid phases and important descriptors like open-circuit voltage and specific capacity to assist in the theoretical characterization of anode materials. Our objective is to provide valuable insights into the properties of Li and Na atoms adsorbed on a stable monolayer of 1T'-MoS$_2$, compared to 2H-MoS$_2$. Our results confirm that the 1T' phase has superior thermal stability to the 2H phase and can also work as an anode material for Na ions.

\section{Computational Methodology}

Within the density functional theory framework, we use the plane-wave self-consistent approach as 
implemented in the Vienna ab-initio simulation program (VASP)\cite{VASP0,VASP1,VASP2}.  
We use the generalized gradient approximation (GGA) devenloped by Perdew-Burke-Ernzerhof (PBE)\cite{PBE} 
and the van der Waals correction through the DFT-D3 with BJ-damping 
correction\cite{grimme2010consistent,grimme2011effect}. 
The kinetic energy cutoff of the wave functions is set to 520 eV. 
In unit cell calculations, we use a dense $21 \times 21 \times 1$ k-grid. For Li/Na adsorption calculations, 
we expand the system to a ${4}\times {4} \times {1}$ supercell for the 2H-phase 
and to a ${2} \times {4} \times {1}$ supercell for the 1T'-phase. 
Note that both supercells (2H and 1T') contain 16 Mo and 32 S atoms. 
Here the integration k-mesh is set to ${5} \times {5}\times {1}$. 
Finally, we perform additional charge analyses by post-processing the charge 
density data using the Bader charge analysis developed by the Henkelman 
group\cite{tang2009grid,sanville2007improved,yu2011accurate}.

We calculated the average absorption energy ($E_{ads}$) of N $\xi$-atoms 
($\xi$ refers to lithium or sodium ad-atoms) in the MoS$_2$ supercell using:
\begin{equation} \label{eq:Eads} 
E_{ads}= \frac{E_{\xi{_N}MoS_2} - E_{MoS_2} -  N\, E_{\xi} } {N},
\end{equation}
where $E_{MoS_2}$ refers to the total energy of pristine MoS$_{2}$ supercell, 
$E_{\xi{_N}MoS_2}$ is the energy of the system with N ad-atoms, corresponding 
to a concentration $x=N/32$ of lithium/sodium ions adsorbed on the MoS$_{2}$ monolayer, 
and $ E_{\xi}$ refers to the chemical potential of a lithium or sodium ion. 
%
Usually, the value of $ E_{\xi}$ is obtained by considering one Li/Na atom in a BCC crystal. 
However, depending on the process, the diatomic gas configuration\cite{de2021dft,pajtler2023lithium} could be used.
Note that in the definition, Eq. \ref{eq:Eads}, more negative values represent stronger adsorptions.

\textcolor{black}{To account for the effects of repulsion between the ions, we calculate the differential adsorption energy\cite{deb2022two} defined as  }
\begin{equation}
E_{\Delta ads}=\frac{E_{\xi{_M}MoS_2}-E_{\xi{_{N}}MoS_2}-(M-N)E_{\xi}}{M-N},
\end{equation}
\textcolor{black}{where $E_{\xi{_{M/N}}MoS_2}$ refers to the total energy associated with $M/N$ adsorbed ions (with $M>N$),  
$ E_{\xi}$ refers to the chemical potential of a lithium or sodium ion in a BCC crystal, and $\xi$ refers to lithium or sodium ions.}

To evaluate the potential of alkali metal-ion batteries, one factor that is often considered 
is the open-circuit voltage (OCV) profile. This open-circuit voltage is linked to 
the variation in chemical potential of the electrode when ion are adsorbed. 
Additionally, it can be associated with the formation energy ($E_{form}=N E_{ads}$) slope 
in relation to the concentration of ions. 
The OCV profile for ion adsorption on a surface can be calculated by measuring the change 
in Gibbs free energy of the system. By ignoring changes in volume and entropy, the OCV 
expression can be simplified to the difference in total energies\cite{akgencc2019two,Ya_Meng_Li_2020,zhou2004first}, as	
\begin{equation} \label{eq:OCV} 
OCV \approx \frac{E_{\xi_{x_2}MoS_2} - E_{\xi_{x_1}MoS_2} - \left(x_2 - x_1 \right)E_{\xi}  } {\left(x_2 - x_1 \right)e},
\end{equation}
where $e$ is the electron charge, $E_{\xi_{x_i}MoS_2}$ is the total energy of the 
MoS$_{2}$ with a concentration of $x_i$ ions, and $E_{\xi}$ is the chemical potential 
of lithium/sodium ion.

When evaluating a material as a potential battery electrode, a key factor to consider 
is its specific capacity, also known as the theoretical gravimetric reversible capacity. 
This refers to the amount of charge that can be stored in the material per unit of mass, 
and is determined by the material's chemical composition.\cite{kim2020applications,gonzalez2022v} 
The specific capacity reads as:  
\begin{equation} \label{eq:capacity} 
C = \frac{  N F } {M_w},
\end{equation} 
where, $N$ is the number of ions, $F$ is the Faraday constant, and $M_w$ is the molecular weight of the electrode.

\textcolor{black}{Another important variable to evaluate the battery performance is the consideration of the performance of a system in the charge/discharge processes, which can induce changes in the volume of the material. The cyclic stability can be associated with the deformation during the lithiation/delithiation or sodiation/desodiation process can lead to a diminution of reversible capacity in a few cycles\cite{deb2022two}. To quantify this induced deformation, we calculate the strain induced and change in buckling thickness upon adsorption of alkali-metal ions. The percentual lattice strain-induced expansion is calculated as }
\begin{equation}
\Delta Z =\frac{a-a_{0}}{a_{0}}\times 100 , \label{eq.Z}
\end{equation}
\textcolor{black}{where $a$ and $a_{0}$ are the lattice constants after and before alkali-metal adsorption.}

To evaluate the thermal stability of our systems, we perform Nose-Parrinello-Rahman ab-initio molecular dynamics calculations using the SIESTA package\cite{soler2002siesta} to evaluate the thermal stability of the structures. We use a double-$\zeta$ polarized basis set, with a Monkhorst-Pack grid of $3\times 3\times 1$. We use the optB88-vdW exchange-correlation functional\cite{klimevs2009chemical}  to incorporate vdW non-local interactions. The simulation runs at constant temperatures of 300 and 400 K for 5 ps with a time step of 1 fs.

\begin{figure*}[t!]
	\centering
	\includegraphics[clip,width=\columnwidth,angle=0]{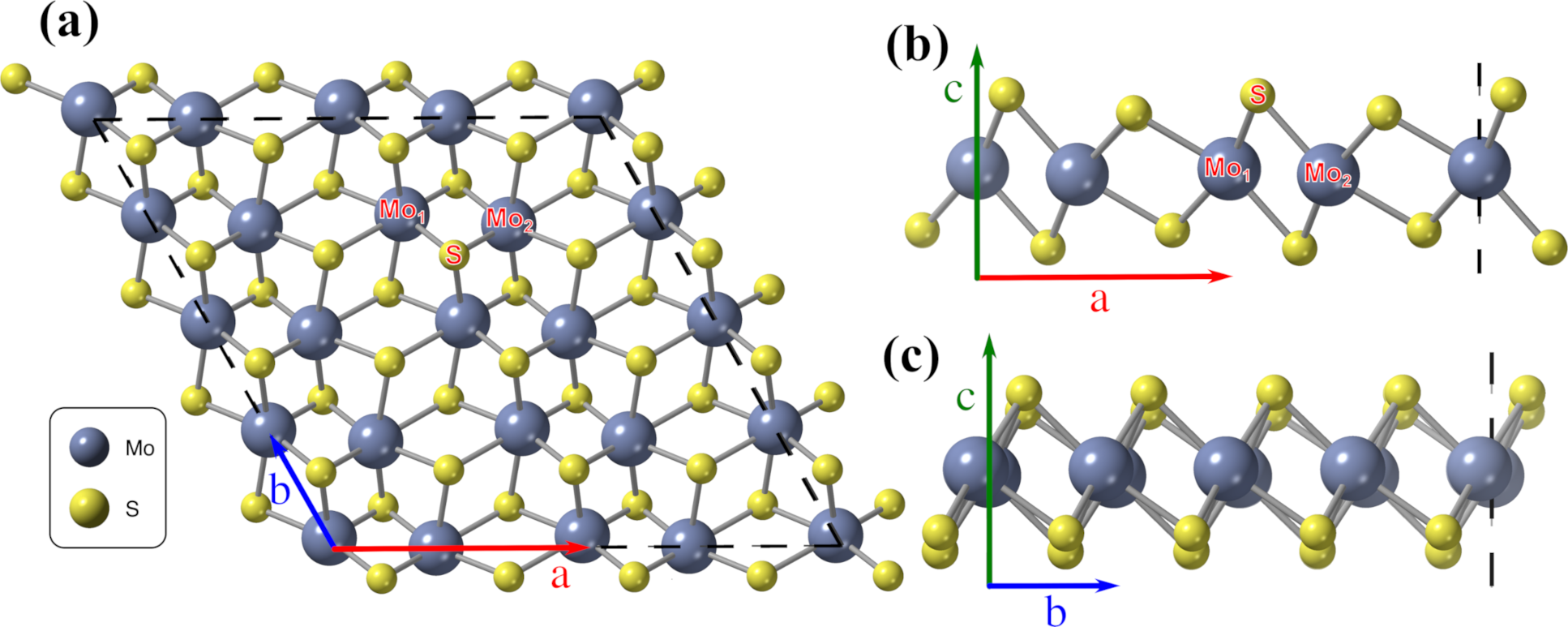} 
	\caption{(Color on-line) 
		Ball-and-stick  representation for the pristine $2\times4\times1$ 
		1T'-MoS$_2$ monolayer supercell: (a) top view, (b) side view along 
		b-axis, and (c) side view along a-axis. \textcolor{black}{The labels Mo$_1$, Mo$_2$ and S indicate the stable top-position adsorption sites.}}
	\label{Fig:scheme}
\end{figure*}

\section{Results and discussion}

The 1T'-MoS$_2$ phase is a distorted structure that comes from the 1T phase. In TMDCs, 
the primary mechanism for forming the 1T' phase is Peierls distortion. The primitive 1T' 
cell is equivalent to a ($1 \times 2 $) supercell of the 1T 
phase\cite{chen2021diverse}. Fig. \ref{Fig:scheme} shows the geometric structure of 
1T'-MoS$_2$ in detail. 
When comparing the total energies of the pristine cells (considering cells with the same number of atoms), 
we find that the system with 2H symmetry is the most stable configuration. 
The 1T' configuration is +9.38 eV (+0.59 eV per Mo atom) above the 2H configuration, 
making it the second most stable configuration. Among the three configurations explored, 
the 1T configuration is the least stable, with +13.44 eV (+0.84 eV per Mo atom) above the 2H phase. 
This result is consistent with previous results\cite{sandoval1991raman,huang2018first}.

\textcolor{black}{To  analyze the possible magnetic character of the different crystalline phases of MoS$_2$ monolayers, we  used crystal field theory to predict how the 4d states of the Mo atom will split. The valence state of the Mo atom in MoS$_2$ is +4, where two d-electrons remain in the outermost layer\cite{zhao2018metastable,xu2018magnetism}. In the 2H phase, the Mo atom is centered in a trigonal prismatic environment (D$_{3h}$) where the z$^2$ state is the lowest energy state. In the 1T phase, Mo is centered in an octahedral structure (O$_h$) where the t$_{2g}$ levels are the lowest in energy. The 1T' phase corresponds to a distortion of the octahedral structure due to the dimerization of the molybdenum atoms; in this case, the xy and x$^2$-y$^2$ orbitals decrease in energy (concerning 1T) because the Jahn-Teller effect reduces the density along the z-axis\cite{zhao2018metastable}.
According to Hunt's rules and the minimum energy principle, we can predict that the magnetic moment of the 1T configuration is 2 $\mu_B$, and 0 $\mu_B$ for the 2H and 1T' configurations. Later, we will analyze the behavior when ions fully saturate the monolayer.}

\begin{figure}[t!]
	\centering
	\includegraphics[clip,width=0.5\columnwidth,angle=0]{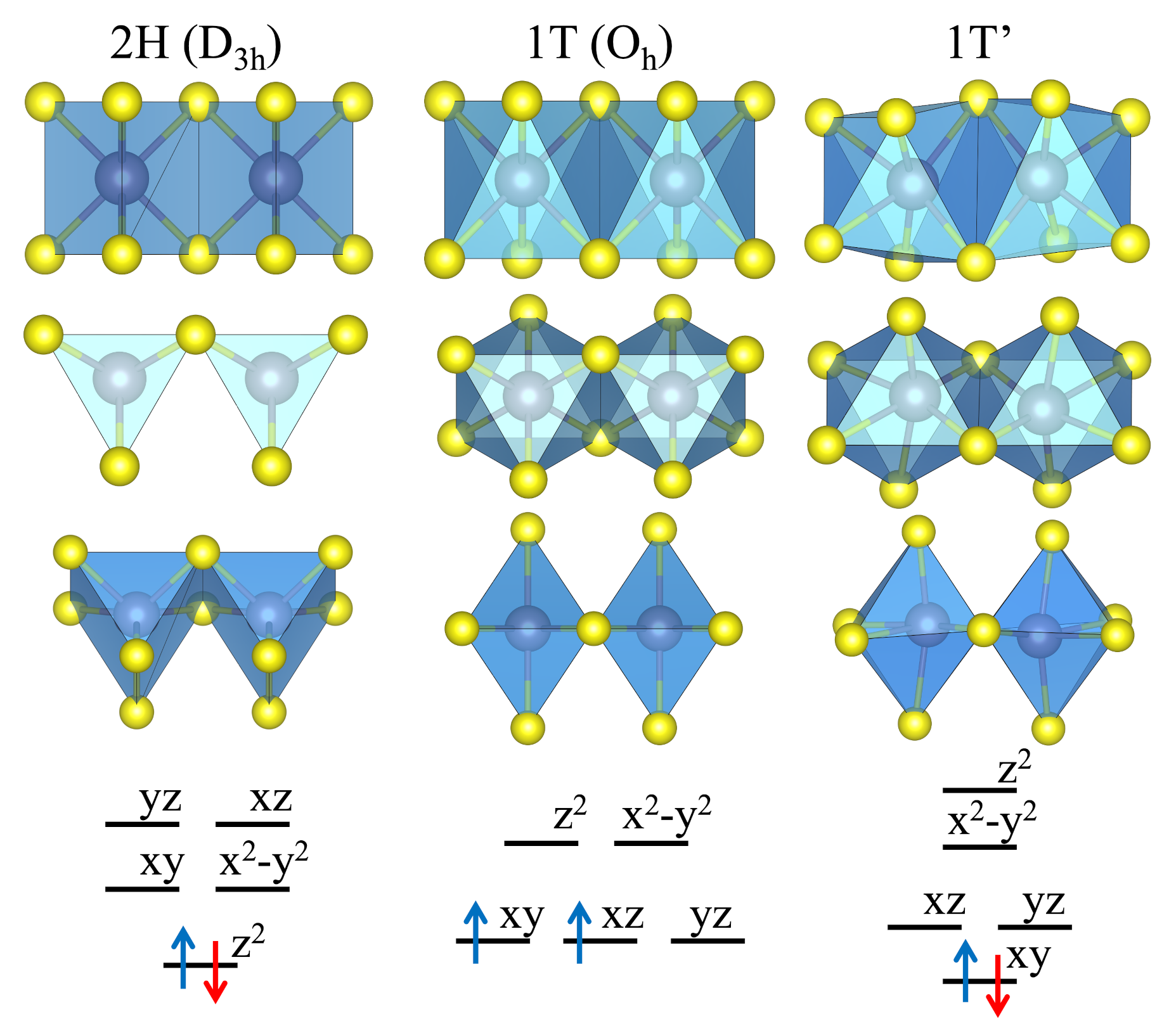} 
	\caption{(Color on-line) \textcolor{black}{The crystallographic environment surrounding Mo in the different crystallographic phases of MoS$_2$ showing a front, top, and side views. Additionally, in the bottom panel, we show the occupation of electrons in Mo 4d orbits for pristine 2H, 1T, and  1T'  MoS$_2$ phases.}}
	\label{Fig:magnetic}
\end{figure}

\subsection{Lithium and Sodium adsorption}

We analyze the adsorption of individual Li and Na atoms onto a supercell of MoS$_2$-1 T' and MoS$_2$-2H containing the same number of Mo and S atoms. The most favorable adsorption site are  determined  by comparing the adsorption energies at high-symmetry sites while considering the ion energy in a diatomic gas and BBC crystal. Our analysis has identified four non-equivalent positions in the 1T' cell, two for top-Mo positions (Mo$_1$, Mo$_2$) and two for top-S positions. However, both top-S configurations yield the same adsorption energy, so we labeled this position S, as shown in Fig. \ref{Fig:scheme}. For the 2H phase, we only had a top-Mo, a top-S, and the hollow sites. The results are presented in Table \ref{Tab:1}. 

\begin{table}
	\begin{tabular}{ccccc}
		\hline
		& \multicolumn{2}{c}{1T{'} E$_{ads}$ (eV) } &\multicolumn{2}{c}{2H E$_{ads}$(eV)}\\
		\hline
		Position  &  Li  & Na   &  Li  & Na    \\
		\hline
		Mo-1 &  -2.06 (-3.23)& -2.06 (-2.76)   & -1.10 (-2.27)  & -1.10 (-1.80)  \\
		Mo-2 &  -1.87 (-3.04)& -1.90 (-2.60)   &               &               \\
		S    &  -1.80 (-2.97)& -1.91 (-2.61)   & -0.41 (-1.57)  & -0.73 (-1.43)  \\  
		Hollow&             &                  & -0.95 (-2.11)  & -1.08 (-1.78)  \\
		\hline
	\end{tabular}
	\caption{Adsorption energy for Li and Na atoms on MoS$_2$ monolayers considering 1T' and 2H phases. 
	We consider the chemical potential of Li/Na, $E_{\xi}$ (Eq. \ref{eq:Eads}) from a BCC crystal\cite{urban2016computational,liu2021modulating,gonzalez2022v} 
	and gas phase\cite{de2021dft,pajtler2023lithium} (in parenthesis). 
	For adsorption positions, we have considered the most symmetric points (details in  Fig. \ref{Fig:scheme}).}\label{Tab:1}
\end{table}

The adsorption energy calculated with the BCC crystal is consistently smaller than the calculated 
with the gas phase (breaking a metal-metal bond in a BCC crystal is about two times more difficult as 
breaking an ion-ion bond in Li$_2$ or Na$_2$). 
In the 1T' phase, the adsorption of  Li/Na occurs at the top-Mo$_1$ positions. Note that the role of Mo$_1$ and Mo$_2$ changes depending on the viewpoint; the labels in Fig. \ref{Fig:scheme} correspond to a top view from $+$z.
In the 2H phase, the top-Mo position is the most favorable adsorption site. In contrast, in the 1T-MoS$_2$ configuration, 
the alkali-metal ions are adsorbed in a bridge position\cite{he20221t}. 
\textcolor{black}{In the low-concentration regime, the distribution of ions has a significant impact on the total energy and adsorption energy. Analyzing the convex hull plot (Fig. S3) at the limit of dilute ions on 1T'-MoS$_2$, we find the following energy-minimizing distribution rules: (1) in this regime, lithium and sodium ions follow the same distribution, (2) ions should be located in top-Mo$_1$ positions, (3) ions should be homogeneously distributed on both faces, and (4) ions should be distributed maximizing the distance between ions on the same face. Following rules (1) and (2), we find variations of $\sim 0.15$ eV in the formation energy. Some relevant configurations are shown in Figs. S1  and  S2  of supplementary material.  }

\begin{table}
	\begin{tabular}{c|c|cc|cc|cc}
		\hline
		Ions & C (mAh/g) &\multicolumn{2}{c|}{ E$_{ads-BCC}$ (eV)} &\multicolumn{2}{c|}{ E$_{\Delta ads-BCC}$ (eV) }& \multicolumn{2}{c}{ $\Delta Z$ (\%)}      \\
		\hline
		   &        & Li     & Na      & Li     & Na     &  Li   &  Na \\     
2	&20.93	&-1.921	&-1.885	&-1.921	&-1.885	&0.09	&0.14 \\
8	&83.70	&-1.603	&-1.234	&-1.497	&-1.017	&0.54	&0.87 \\
16	&167.41	&-1.153	&-0.844	&-0.703	&-0.454	&1.24	&1.56 \\
32	&334.81	&-0.665	&-0.455	&\textbf{-0.176}	&-0.066	&0.93	&1.45 \\
48	&502.22	&-0.377	&-0.314	&0.198	&\textbf{-0.031}	&0.57	&2.02 \\
64	&669.62	&-0.276	&-0.210	&0.027	&0.103	&-0.11	&1.78 \\
		\hline
	\end{tabular}
	\caption{\textcolor{black}{Specific capacity (C), adsorption energy E$_{ads-BCC}$ and differential adsorption energy E$_{\Delta ads-BCC}$ for Li and Na ions on 1T'-MoS$_2$ monolayer. As a reference, we use the energy of Li/Na in BCC crystal. The adsorbed ions vary from 2 (concentration 0.13) to 64 (concentration 4.00). }}\label{Tab:2new}
\end{table}

\textcolor{black}{Table \ref{Tab:2new} shows the specific capacity, adsorption, and differential adsorption energy for several ions adsorbed on 1T'-MoS$_2$ monolayer. The specific capacity linearly increases with the number of ions in the cell (Eq. \ref{eq:capacity}). Both adsorption energies increase as the ion concentration increases, tending to positive values. 
In our definition, the negative adsorption energies indicate the system is thermodynamically favorable. Using the differential energy (with lithium and sodium in BCC crystal) as a criterion, we estimate that the 1T'-MoS$_2$ phase has a maximum ion absorption capacity of about 32 lithium (concentration $\sim$2) and 48 sodium (concentration $\sim$3), which is equivalent to a maximum specific capacity of $\sim$335 mAh/g for lithium and $\sim$503 mAh/g for sodium. In this high-concentration regime, lithium forms flat layers, whereas sodium forms staggered planes. Details in Fig. S4.}

\textcolor{black}{The lattice constant variation induced by the ion adsorption is quantified with $\Delta Z$ (defined in Eq. \ref{eq.Z}), shown in last two columns of Table \ref{Tab:2new}. Larger $\Delta Z$  values indicating a significant deformation during the lithiation/delithiation or sodiation/desodiation process can lead to a diminution of reversible capacity in a few cycles. 
Due to its larger atomic radius, Na induces larger lattice strain-induced deformations than Li. The maximum $\Delta Z$ is found for 48 Na with a $\Delta Z$ = 2.0 \%; for Li, the maximum is found for 16 Li, with a $\Delta Z$ = 1.3 \%. The $\Delta Z$ values below 2 \% are consistent with low lattice strain-induced expansion, indicating higher cyclic stability.}

\subsection{Diffusion path}
\begin{figure}[!hb]
	\centering
	\includegraphics[clip,width=0.5\columnwidth,angle=0]{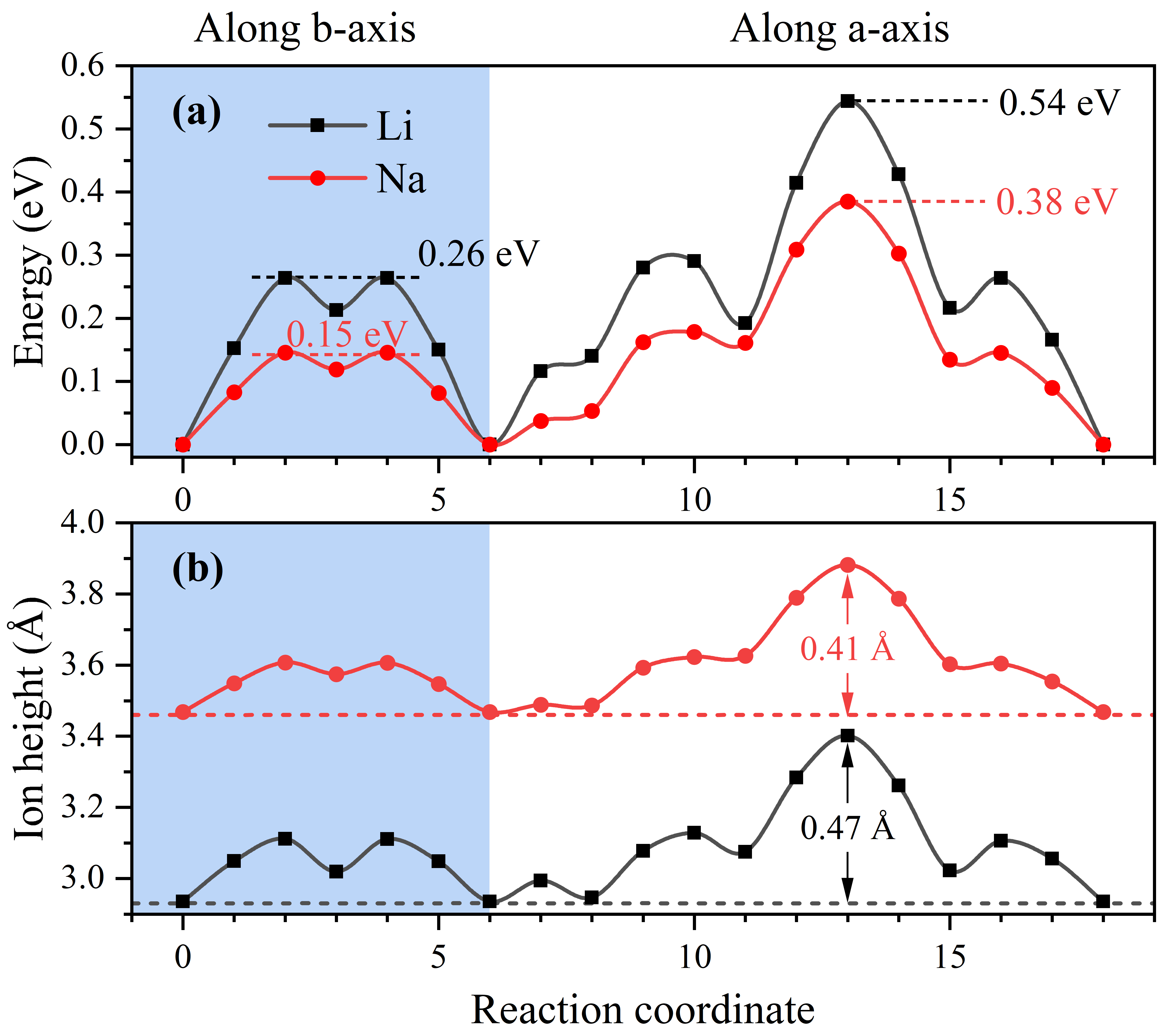} 
	\includegraphics[clip,width=0.5\columnwidth,angle=0]{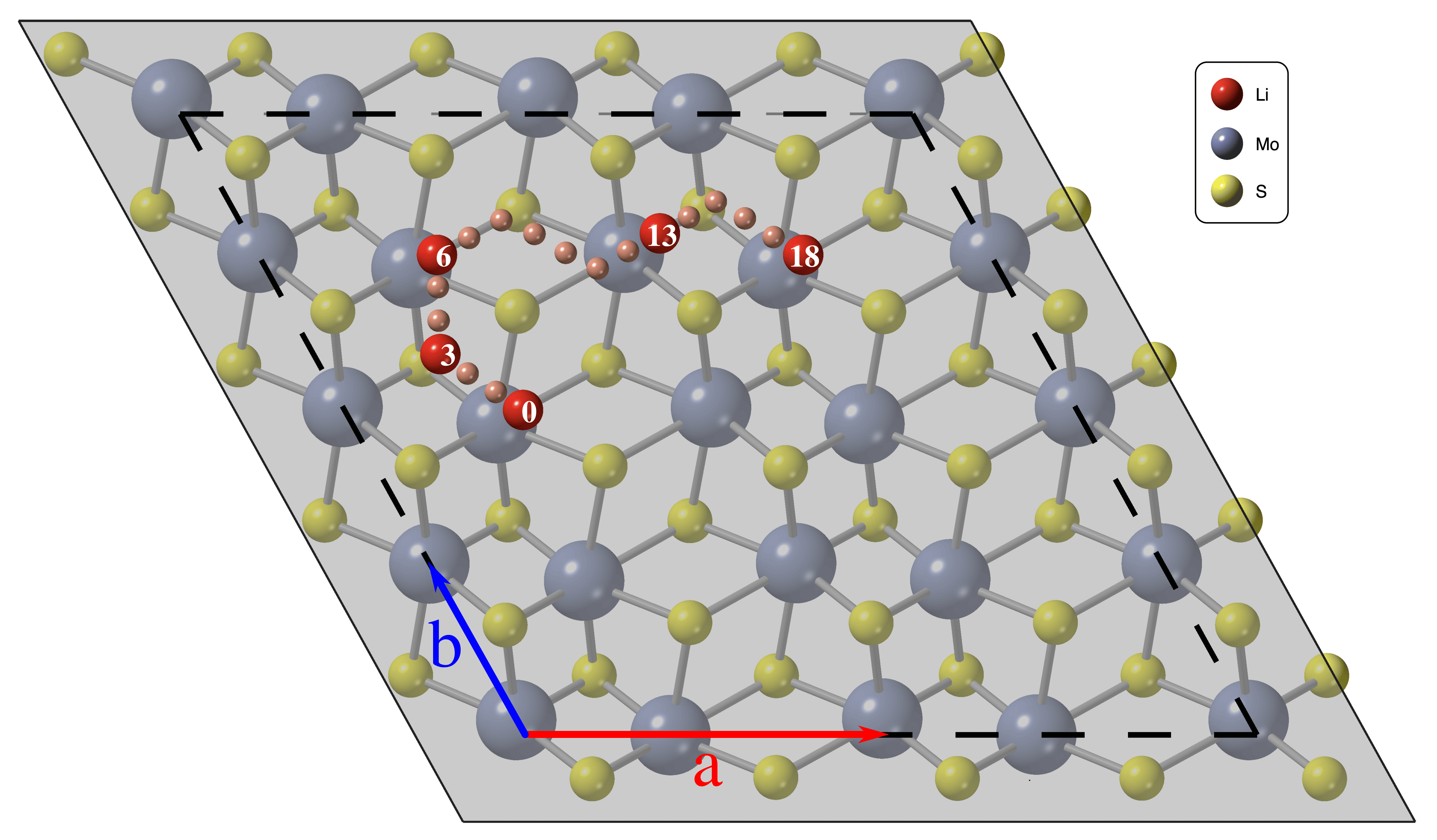} 
	\caption{(Color online) Nudged-elastic band (NEB) minimum energy pathway for the 
	lithium and sodium ion diffusion in a 1T'-MoS$_2$ monolayer. In (a), we present 
	the relative energy and (b) ion-height variation --considering the distance between 
	the ion and the average position on the z-axis of the molybdenum atoms. 
	In the bottom panel, the ball-stick representation of the nudged-elastic-band (NEB) 
	path of lithium diffusing in 1T'-MoS$_2$ monolayer. Note that sodium ions follow 
	the same path as lithium but with a higher height. 
	The smaller pink spheres represent the NEB minimum energy pathway, and the larger 
	red spheres identify the extreme points in (a) and (b) panels. The positions 0, 6, 
	and 18 correspond to top-Mo-2, positions 3 and 15 are top-S, and position 11 is top-Mo-1.
	}
	\label{Fig:neb1T}
\end{figure}

The Nudged-elastic band (NEB) is used to study the diffusion of Li and Na ions on the 1T'-MoS$_2$ monolayer. The NEB method allows us to study kinetic behavior by identifying the energy barriers of the ions moving 
along one path; given the results in Table \ref{Tab:1}, we explore the diffusion of Li/Na moving along the $\vec{a}$ and $\vec{b}$ directions. The NEB path starts and ends at the most stable top-Mo-1 adsorption sites. The diffusion path, energy barriers, and equilibrium height for lithium and sodium on the 1T' surface are depicted in Fig. \ref{Fig:neb1T}. The energy profile along the chosen path (Fig. \ref{Fig:neb1T}(a)) 
reveals that Li and Na atoms present anisotropic diffusion, where the ions easily diffuse along the b-direction. 

Generally, the lithium diffusion barrier in the 1T'-MoS$_2$ monolayer is higher than in sodium, Fig. \ref{Fig:neb1T}(a). Along the b-direction, the diffusion barrier reaches 0.26 eV for Li, whereas, 
for Na, such barrier is lower, 0.15 eV. This stronger interaction between lithium and the surface is also reflected in the equilibrium height (Fig. \ref{Fig:neb1T}(b)), which is lithium is 2.94 \AA{} and in sodium is 3.47 \AA{}. However, we note that the variation in the height of the ions along the entire path is almost the same for Li/Na, where the sodium ion is 0.6 \AA{} above its lithium counterpart. The barriers we found are lower than the Li and Na atoms on 1T-MoS$_2$\cite{he20221t} and graphite\cite{persson2010thermodynamic}. However,  the energy diffusion barrier is highest than the  0.16 eV for Li and 0.13 eV for Na found on the A'-MoS$_2$ surface\cite{sukhanova2022novel}. The studied diffusion of ions in 2H-MoS$_2$ is present in Fig. S3. Due to the symmetry of the cell, the diffusion along the  $\vec{a}$ and $\vec{b}$ axis is the same. Starting from a Top-Mo position, the activation barrier for lithium is similar to the previous case, with a value of 0.23 eV; for sodium, the barrier decreases dramatically, reaching 0.08 eV. The higher barrier configuration corresponds to a bridge-like position with an ion height of just 0.1 \AA{} above the Top-Mo ground state. The results also show the 1T'-MoS$_2$ monolayer is higher than in sodium1T'-MoS$_2$ monolayer is higher than in sodium, Fig. \ref{Fig:neb1T}(a). Along the b-direction, the diffusion barrier reaches 0.26 eV for Li, whereas, for Na, such barrier is lower, 0.15 eV. This stronger interaction between lithium and the surface is also reflected in the equilibrium height (Fig. \ref{Fig:neb1T}(b)), which is lithium is 2.94 \AA{} and in sodium is 3.47 \AA{}. However, we note that the variation in the height of the ions along the entire path is almost the same for Li/Na, where the sodium ion is 0.6 \AA{} above its lithium counterpart. The barriers we found are lower than the Li and Na atoms on 1T-MoS$_2$\cite{he20221t} and graphite\cite{persson2010thermodynamic}. However,  energy diffusion barrier is highest than the  0.16 eV for Li and 0.13 eV for Na found on the A'-MoS$_2$ surface\cite{sukhanova2022novel}. The studied diffusion of ions in 2H-MoS$_2$ is present in Fig. S3. Due to the symmetry of the cell, the diffusion along the  $\vec{a}$ and $\vec{b}$ axis is the same. Starting from a Top-Mo position, the activation barrier for lithium is similar to the previous case, with a value of 0.23 eV; for sodium, the barrier decreases dramatically, reaching 0.08 eV. The higher barrier configuration corresponds to a bridge-like position with an ion height of just 0.1 \AA{} above the Top-Mo ground state. The results also show both ions in a top-S metastable state, with the ions in a hollow position. 

\subsection{Ions  concentration effect and  battery performance}

\begin{figure}[!ht]
	\centering
	\includegraphics[clip,width=0.5\columnwidth,angle=0]{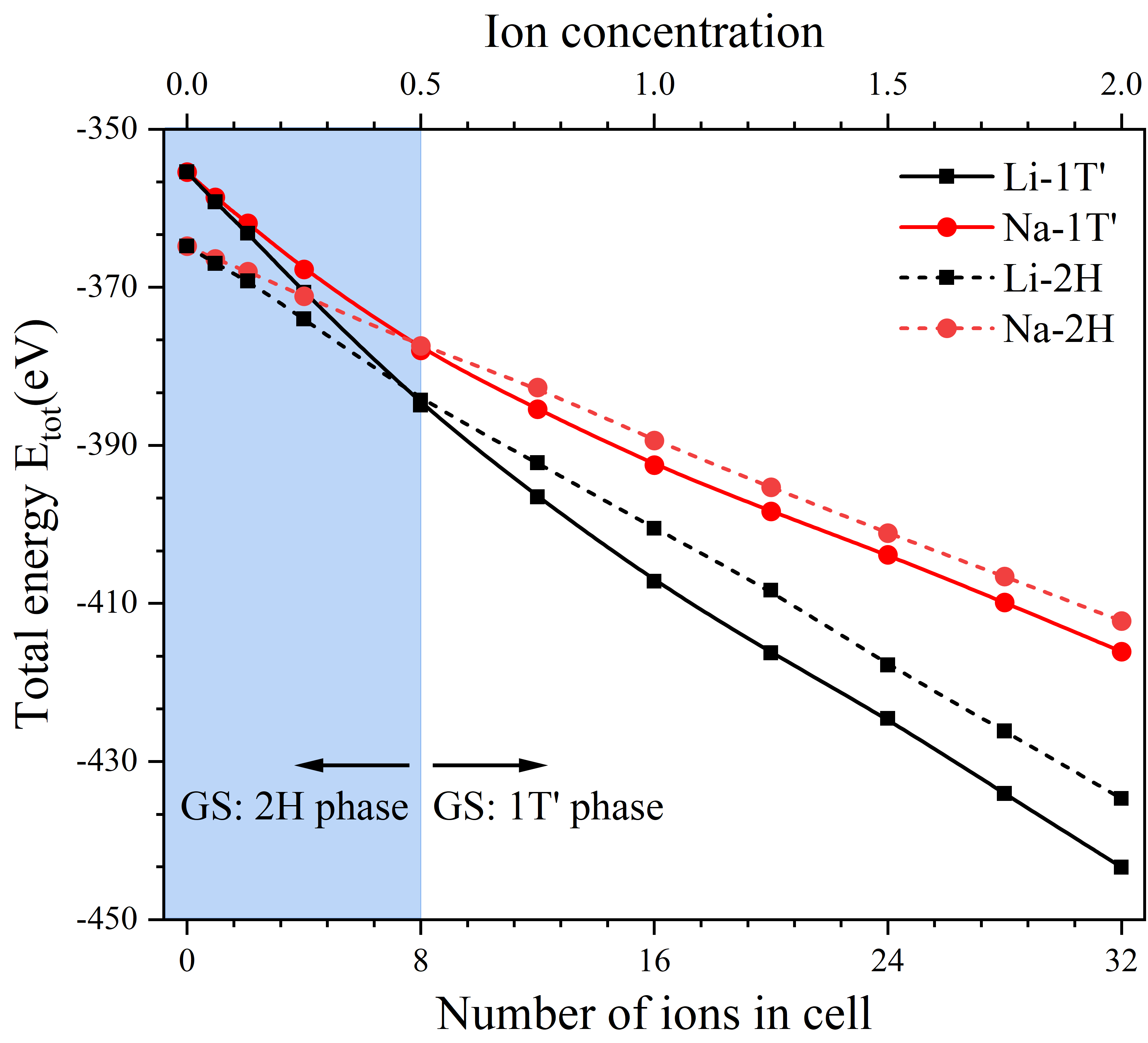} 
	\caption{(Color on-line) Total energy as a function of atoms concentration 
	for Li and Na ions adsorbed on 1T' and 2H MoS$_2$ monolayer. The ion concentration refers to the average 
	density of adsorbed ions per Mo atom, where a concentration of 2.00, means that each Mo in the supercell 
	is bonded with two Li (or Na). The number of Li/Na, Mo, and S in each configuration is the same, 
	allowing us to compare the total energies directly.   
	}\label{Fig:etot-dist}
\end{figure}

When two systems have the same number of atoms, the total energy allows to identify 
changes in the order of stability between different crystalline phases; in this case,  
the total energy is  analyzed as the number of adsorbed ions changes. The concentration of ions in the supercell is determined by the number of alkali ions present over Mo-sites. This concentration can vary from 0.00 to 2.00, as each Mo atom can be bonded with two alkali atoms (one ion on each layer face). It is possible to achieve concentrations above 2.00 by taking into account multiple ion layers\cite{he20221t}. However, the ion-MoS$_2$ interaction is no longer the dominant factor and is instead replaced by an ion-ion interaction. 

In the evolution of the total energy of the MoS$_2$ with a concentration of adsorbed-ions 
(Fig. \ref{Fig:etot-dist}), are identified two regimes. On the one hand, a low-concentration regime where ion concentration is lower than 0.50  (equivalent to a Mo-occupancy of 25 \%), the dashed-line corresponding to the 2H system is below the solid-line, indicating that the 2H-MoS$_2$ configuration is more favorable than the 1T'. On the other hand, for concentrations higher than 0.50, the solid-line, corresponding to the 1T' system, is below; this could indicate a possible change in the stability order, favoring the 1T'-MoS$_2$ over the 2H. For concentrations equal to 0.50, both crystallographic phases are equally probable. 

\begin{figure}[h!]
	\centering
	\includegraphics[clip,width=0.5\columnwidth,angle=0]{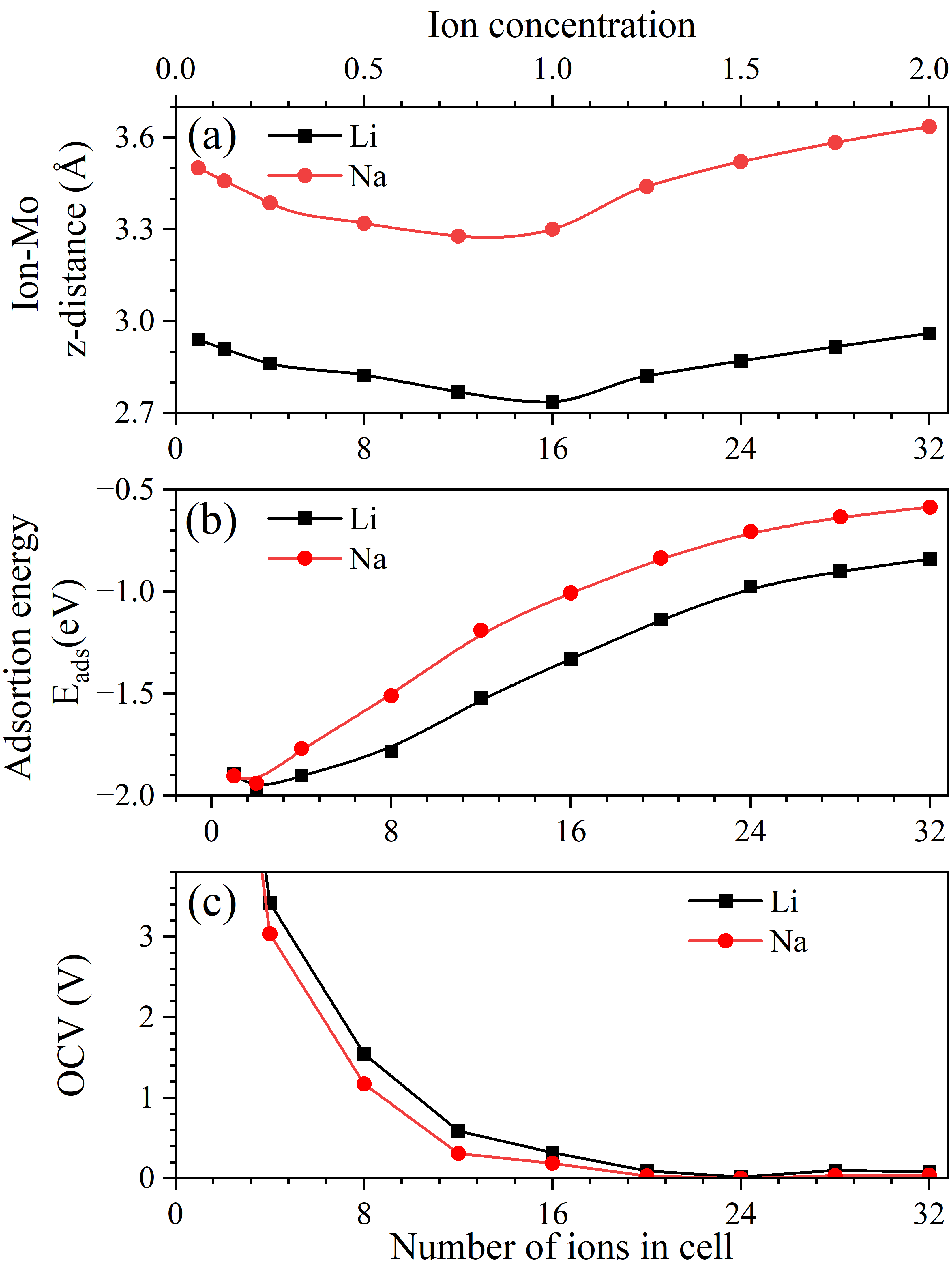} 
	\caption{(Color on-line)
		Characterization of the monolayer 1T'-MoS$_2$ as a function of the concentration of adsorbed alkali ions. 
		In (a), we present the average ion-surface distance; in (b), we show the adsorption energy; 
		and in (c), we present the open circuit voltage (OCV). The lower horizontal axis represents the 
		ion concentration in the supercell, and the upper horizontal axis represents the theoretical specific capacity (Eq. \ref{eq:capacity}).
		The black line corresponds to lithium and the red line to sodium.}\label{Fig:many}
\end{figure}

Once we consider that the relevant polymorph for application as an ion-alkali battery electrode corresponds to the 1T'-MoS$_2$ phase, let us study how the physical properties change when we vary the ion concentration. First, in Fig. \ref{Fig:many}(a), we show the average ion-surface distance changes depending on the ion concentration on 1T' phase. Two trends are observed: for ion-concentrations $\leq 1.00$, the ion-surface distance decreases as we increase the number of ions, while for concentrations $> 1.00$, the distance increases with the number of ions. In the next section, we will associate this behavior to charge transfer between the ion and the surface, where the concentration equal to 1.00 becomes an inflection point for the charge-transfer curve.

In the case of sodium, the average ion-surface distance changes between 3.3 \AA{} and 3.6 \AA{}, 
while for lithium, the ion-surface distance is smaller, varying between 2.95 \AA{} and 2.72 \AA{}. 
This is also reflected in the adsorption energy (Fig. \ref{Fig:many}(b)), where Na has lower adsorption energy than Li, such difference is sustained as the concentration increases. For concentrations above 0.50, the difference in adsorption energies for Li and Na remains almost constant. The larger ion-surface height and the lower adsorption energy suggest that Na has more mobility on the 1T'-MoS$_2$ surface.

To check thermally stability of the fully saturated systems (ion concentration of 2.00), we perform ab-initio molecular dynamics (MD) using the SIESTA package\cite{artacho2008siesta}. We consider supercells with the same number of Mo sites in 1T' and 2H phases (16 Mo sites). We simulate the temporal evolution of the systems at constant temperatures of 300 K  and 400 K with a time step of  1 fs for a maximum time of 5 ps for  1T'-MoS$_2$ monolayer. We consider Full-ion coverage (ion concentration of 2.00). Detailed information is included in the supplementary material (Figs. S3 to S6).  
For the 2H-MoS$_2$, the adsorption of Li-ions became unstable after  0.6 ps at  300 K, and it is fully dissociated after 0.3 ps at 400 K. This behavior was not observed after 3 ps for  Na ions on  2H  phases. Na ions tend to rearrange, forming multiple  Na-layers\cite{he20221t}; we speculate that this is due to Na larger ionic radius, further analysis in this direction is beyond the scope of the current work.
Our MD calculations indicate that the 2H-MoS$_2$ phase became thermally unstable when fully covered by Li ions. These results align with previous experimental studies that show Li ions induce phase transitions for 2H-MoS$_2$\cite{hou2022phase}. 
In contrast, the molecular dynamics in the 1T'-MoS$_2$ phase of the system fully saturated by lithium and sodium ions are stable at 300 K and 400 K.

By examining Fig. \ref{Fig:many} (a, b), we can see the effect of increasing ion concentration on the adsorption properties of alkali ions on MoS$_2$, which in turn affects their performance to storage energy.  To measure this impact, we assess the battery's specific capacity and open circuit voltage (OCV) as parameters to evaluate the battery capabilities\cite{he20221t}. 
Regarding theoretical specific capacity, Na-ion batteries with TiO$_2$ nanotube as anode material\cite{xiong2011amorphous}  have a specific capacity of 150 mAh/g. Experimental reports also show that a combination of MoS$_2$-RGO can reach specific capacities of 305 mAh/g\cite{qin2015mos2}. Previous theoretical calculations estimate specific capacities of 1172 mAh/g and 335 mAh/g for Li and Na on 1T-MoS$_2$, respectively. However, for Li, they consider 
up to seven layers of Li ions on each side of the monolayer\cite{he20221t}. 

In contrast, as we only consider a layer of alkali ions on each side of the MoS$_2$ monolayer, for Li and Na ions, we find a maximum specific theoretical capacity on MoS$_2$ monolayer of 335 mAh/g. 
In Fig. \ref{Fig:many}, the lower horizontal axis corresponds to the ion concentration, and the upper horizontal axis corresponds to the theoretical specific capacity; in the Eq. \ref{eq:capacity}, we can identify a linear relationship between both parameters.

In Fig. \ref{Fig:many}(c), the open circuit voltage profile for Li/Na follows the same trend, where the open circuit voltage decreases with increasing ion content. As the Li curve is above the Na, the energy storage of lithium on 1T'-MoS$_2$ is better than the Na counterpart.

When the ion concentration is below 1.00, both alkali-ion batteries can deliver an average output voltage above 2.5 V. This is significantly higher than commercial lead-acid batteries ($\sim$2.0 V) and rechargeable alkali manganese batteries ($\sim$1.5 V)\cite{liu2021modulating}. However, the average output voltage for concentrations between 1.00 and 2.00 is 0.67 V for lithium and 0.39 V for sodium. This indicates a worse performance in the latter case. The OCV results suggest that the optimal operating Li/Na ion concentration for the 1T'-MoS$_2$ monolayer is below 1.00, 
which is equivalent to a maximum theoretical specific capacity of 160 mAh/g.

%
%

\subsection{Electronic  structure analysis}

\begin{figure}[t!]
	\centering
	\includegraphics[clip,width=0.5\columnwidth,angle=0]{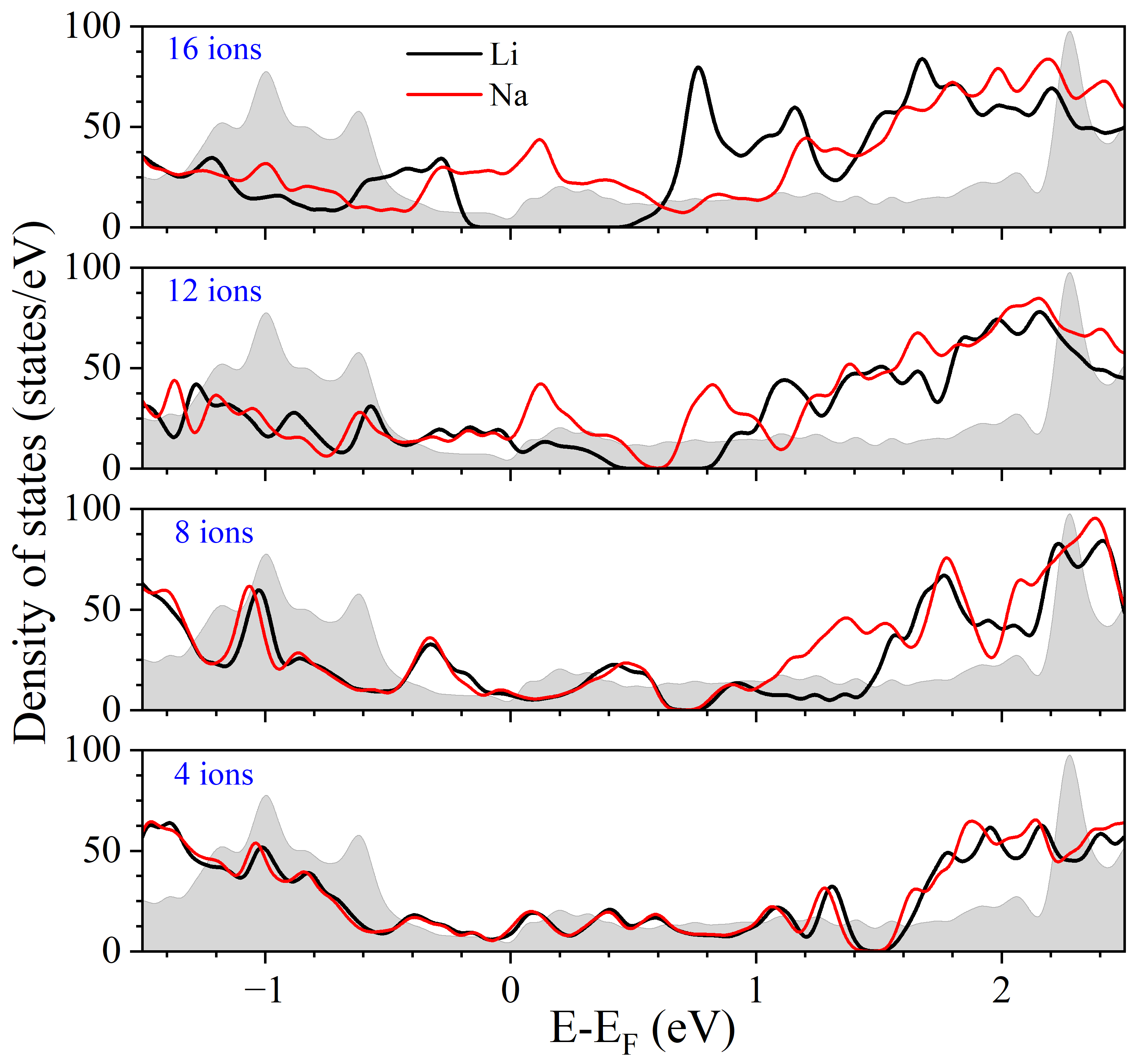} 
	\caption{The density of states (DOS) for the 1T'-MoS$_2$ for several alkali-ion concentrations: 0.25 (4 ions), 0.50 (8 ions), 0.75 (12 ions), and 1.00 (16 ions). The black line corresponds to adsorbed lithium, the red line corresponds to adsorbed sodium, and the shadow region corresponds to the pristine monolayer 1T'-MoS$_2$. }
	\label{Fig:dos_1Tp}
\end{figure}

Finally, we correlate the properties described in the previous sections with modifications in the electronic properties of the system and charge transfer effects as the ion concentration increases.
We characterize the electronic structure through the density of states. For the 1T'-MoS$_2$ monolayer, we change the number of ions adsorbed on the surface, considering 0, 4, 8, 12, and 16 ions. In Fig. \ref{Fig:dos_1Tp}, the shaded region (the same for all panels) represents the density of states of the pristine 1T'-MoS$_2$ monolayer with a metallic character.
The black line represents the density of states in the presence of adsorbed lithium ions, and the red line represents the density of states with sodium ions. When the ion concentration is below 0.50 (equivalent to 8 ions in the $2 \times 4$ 1T'-MoS$_2$ supercell), the density of states for both ions (Li/Na) are similar.
 
At a 0.25 ion concentration (4 ions in supercell), a band gap of approximately $\sim$ 0.2 eV opens around an energy of 1.5 eV. When the concentration reaches 0.50 (8 ions in supercell), although the magnitude of the band gap is similar, the position is shifted to 0.7 eV. 
For 12 lithium ions in supercell (0.75 ion concentration), a band gap of 0.4 eV appears centered around 0.6 eV. In contrast, for the same concentration of sodium ions, the band gap appears around the same energy, but the magnitude of the gap is smaller, 0.7 eV. 
When the ion concentration reaches 1.00, indicating half the supercell capacity, the system with 16 lithium ions is a semiconductor with a band gap of 0.8 eV, while its sodium counterpart is now metallic.

In the supplemental material, we provide the density of states for 2H monolayer with 0, 4, 8, 12, and 16 ions adsorbed. The electronic structure of the 2H monolayer shows that it is a semiconductor with a band gap of 1.7 eV. However, when the concentration of alkali ions 
increases to 0.25 per supercell (4 ions), the system becomes metallic, and the band gap decreases to 1.0 eV around an energy of -0.8 eV. The same scenario occurs when the concentration reaches 0.50 (8 ions), where the energy gap decreases further to 0.4 eV. For low concentration 
regimens, the density of states increases near the Fermi level for both Li and Na ions, resulting in an excellent electronic conductor. The reversible control of the band gap between a semiconductor and metallic character using a few alkali ions has potential applications 
for sensor devices.

When analyzing the charge transfer, we find that lithium adsorption on 2H-MoS$_2$ yields a charge transfer to the neighboring atoms, mainly sulfur. This is reflected in the gain of 0.03 e$^-$ by the host molybdenum and 0.20 e$^-$ per atom by each sulfur atom.  However, the charge transfer to the surface is relatively lower when sodium is present, with the host molybdenum gaining only 0.02 e$^-$ and the first neighboring sulfur atoms gaining 0.15 e$^-$ per atom.

\begin{figure}[!hb]
	\centering
	\includegraphics[clip,width=0.5\columnwidth,angle=0]{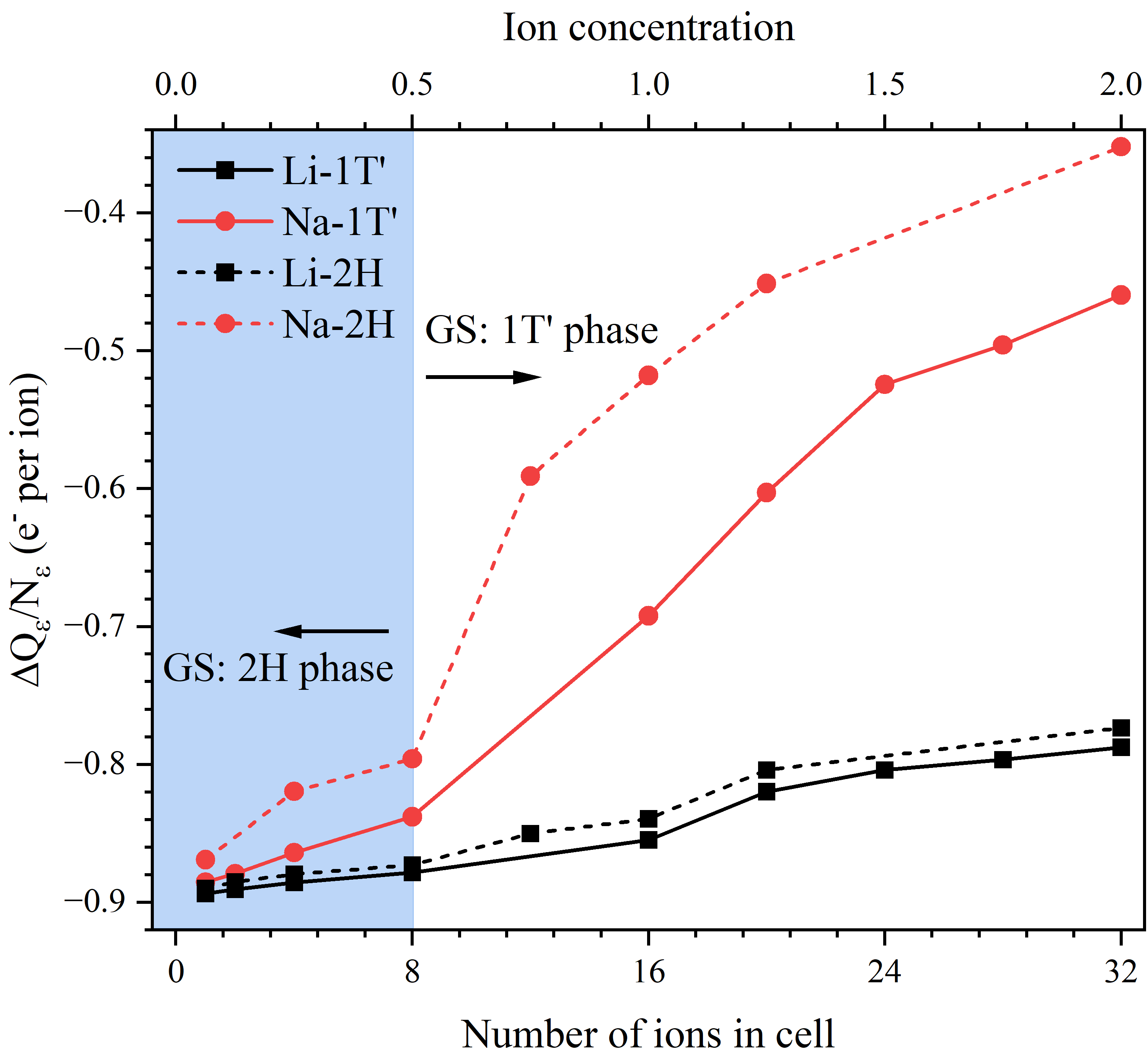} 
	\caption{(Color on-line) Charge transfer between ion and surface based on a Bader charge analysis. Continuous lines correspond to 
	1T'-MoS$_2$ and dashed lines 2H-MoS$_2$. The black line corresponds to adsorbed lithium ions and the red line corresponds to 
	adsorbed sodium ions. }
	\label{Fig:2hbader}
\end{figure}

In Fig. \ref{Fig:2hbader}, we observe how the charge transfer of both types of ions and both surfaces decreases with increasing ion concentration. The adsorption of Na ions results in a lower charge transfer; furthermore, the charge transfer is lower in the 2H phase.
For instance, for the fully saturated configuration (ion concentration equal to 2.00) in the 1T' phase, the transfer reaches -0.45 e$^-$ per ion, and in the 2H phase, it is -0.35 e$^-$ per ion, the same analysis on lithium on both surfaces leads to charge 
transfers in the order of -0.8 e$^-$ per ion.
In the figure, it is also noticeable how for concentrations between 0.50 and 1.00, an inflection point appears in the four curves; 
this transition is linked to the properties observed in Fig. \ref{Fig:many}.

\textcolor{black}{For the 2H and 1T' phases, the effect of the ions (Li and Na) on the different magnetic configurations of MoS$_2$: non-magnetic (NM), ferromagnetic (FM), and antiferromagnetic (AF: Néel, stripy and zig-zag)\cite{kumar2019magnetism}. Pristine and single ion adsorbed systems produce non-magnetic systems. In the limiting case, with 32 Li ions in the cell (concentration 2.00), the high charge transfer of Li (Fig. \ref{Fig:2hbader}) induces ferromagnetism in both MoS$_2$ phases. 
In the 2H-phase, with a total supercell energy of 1.9 meV below the NM configuration, and in the 1T'-phase, the total supercell energy is 43.0 meV below the NM configuration. These results are compatible with the discussion of Fig. \ref{Fig:magnetic}.
Due to the minor charge transfer from Na ions to MoS$_2$ (Fig. \ref{Fig:2hbader}), when we explore the case of 32 Na in the supercell, the system remains non-magnetic (NM).Note that thermal fluctuations make it difficult to observe the small energy differences between the FM and NM configurations.}

\section{Final Remarks} 
Through the first-principles approach that accounts for van der Waals interaction, 
we have investigated the adsorption, diffusion, and storage capabilities of alkali ions 
(specifically Li/Na) in the MoS$_2$ monolayers. Our study primarily focuses on the most 
stable phases (i.e., 2H/1T') in the context of ion batteries. Our ab-initio molecular 
dynamics simulation demonstrates that the 1T' phase remains thermally stable even as 
ion concentration increases, indicating that this phase is well-suited for developing 
alkali ion batteries. Furthermore, our characterization of the performance of 1T'/2H phases 
as batteries through the open circuit voltage (OCV) and specific capacity measurements 
suggests that the 1T'-MoS$_2$ monolayer possesses significant potential for use as cathode 
materials for Li and Na ion batteries, particularly in concentrations below 1.00. These 
findings offer exciting possibilities for the future of ion battery technology and 
demonstrate the significant potential for the practical application of MoS$_2$ monolayers in this field.

\begin{acknowledgement}

JWG would like to acknowledge the financial support from Chilean FONDECYT grants numbers 1221301 and 1220700.
The authors would like to thank the fruitful discussions with Dr. Juliet Aristizabal. 
Additionally, JDC and EF would like to express gratitude 
for the support and use of computational facilities provided by the Laboratorio de Simulaci\'on y 
Computaci\'on Cient\'ifica at Universidad de Medell\'in.

\end{acknowledgement}

\begin{suppinfo}

The following files are available free of charge.
\begin{itemize}
  \item Optimal positions for Li or Na adsorption on 1T'-MoS$_2$.
  \item Convex-hull plot.
  \item Differential adsorption energy for low-concentration regime.
  \item  Multiple adsorbed layers.
  \item NEB and LDOS for 2H-MoS$_2$.
  \item Molecular dynamics information. 
  \item Geometry files.
 and additional information.
\end{itemize}

\end{suppinfo}


\begin{mcitethebibliography}{52}
\providecommand*\natexlab[1]{#1}
\providecommand*\mciteSetBstSublistMode[1]{}
\providecommand*\mciteSetBstMaxWidthForm[2]{}
\providecommand*\mciteBstWouldAddEndPuncttrue
  {\def\EndOfBibitem{\unskip.}}
\providecommand*\mciteBstWouldAddEndPunctfalse
  {\let\EndOfBibitem\relax}
\providecommand*\mciteSetBstMidEndSepPunct[3]{}
\providecommand*\mciteSetBstSublistLabelBeginEnd[3]{}
\providecommand*\EndOfBibitem{}
\mciteSetBstSublistMode{f}
\mciteSetBstMaxWidthForm{subitem}{(\alph{mcitesubitemcount})}
\mciteSetBstSublistLabelBeginEnd
  {\mcitemaxwidthsubitemform\space}
  {\relax}
  {\relax}

\bibitem[Dai \latin{et~al.}(2019)Dai, Liu, and Zhang]{dai2019strain}
Dai,~Z.; Liu,~L.; Zhang,~Z. Strain engineering of 2D materials: issues and
  opportunities at the interface. \emph{Advanced Materials} \textbf{2019},
  \emph{31}, 1805417\relax
\mciteBstWouldAddEndPuncttrue
\mciteSetBstMidEndSepPunct{\mcitedefaultmidpunct}
{\mcitedefaultendpunct}{\mcitedefaultseppunct}\relax
\EndOfBibitem
\bibitem[Zhu \latin{et~al.}(2017)Zhu, Ha, Zhao, Zhou, Huang, Yue, Hu, Sun,
  Wang, Lee, \latin{et~al.} others]{zhu2017recent}
Zhu,~J.; Ha,~E.; Zhao,~G.; Zhou,~Y.; Huang,~D.; Yue,~G.; Hu,~L.; Sun,~N.;
  Wang,~Y.; Lee,~L. Y.~S., \latin{et~al.}  Recent advance in MXenes: A
  promising 2D material for catalysis, sensor and chemical adsorption.
  \emph{Coordination Chemistry Reviews} \textbf{2017}, \emph{352},
  306--327\relax
\mciteBstWouldAddEndPuncttrue
\mciteSetBstMidEndSepPunct{\mcitedefaultmidpunct}
{\mcitedefaultendpunct}{\mcitedefaultseppunct}\relax
\EndOfBibitem
\bibitem[Glavin \latin{et~al.}(2020)Glavin, Rao, Varshney, Bianco, Apte, Roy,
  Ringe, and Ajayan]{glavin2020emerging}
Glavin,~N.~R.; Rao,~R.; Varshney,~V.; Bianco,~E.; Apte,~A.; Roy,~A.; Ringe,~E.;
  Ajayan,~P.~M. Emerging applications of elemental 2D materials. \emph{Advanced
  Materials} \textbf{2020}, \emph{32}, 1904302\relax
\mciteBstWouldAddEndPuncttrue
\mciteSetBstMidEndSepPunct{\mcitedefaultmidpunct}
{\mcitedefaultendpunct}{\mcitedefaultseppunct}\relax
\EndOfBibitem
\bibitem[Gjerding \latin{et~al.}(2021)Gjerding, Taghizadeh, Rasmussen, Ali,
  Bertoldo, Deilmann, Kn{\o}sgaard, Kruse, Larsen, Manti, \latin{et~al.}
  others]{gjerding2021recent}
Gjerding,~M.~N.; Taghizadeh,~A.; Rasmussen,~A.; Ali,~S.; Bertoldo,~F.;
  Deilmann,~T.; Kn{\o}sgaard,~N.~R.; Kruse,~M.; Larsen,~A.~H.; Manti,~S.,
  \latin{et~al.}  Recent progress of the computational 2D materials database
  (C2DB). \emph{2D Materials} \textbf{2021}, \emph{8}, 044002\relax
\mciteBstWouldAddEndPuncttrue
\mciteSetBstMidEndSepPunct{\mcitedefaultmidpunct}
{\mcitedefaultendpunct}{\mcitedefaultseppunct}\relax
\EndOfBibitem
\bibitem[Manzeli \latin{et~al.}(2017)Manzeli, Ovchinnikov, Pasquier, Yazyev,
  and Kis]{manzeli20172d}
Manzeli,~S.; Ovchinnikov,~D.; Pasquier,~D.; Yazyev,~O.~V.; Kis,~A. 2D
  transition metal dichalcogenides. \emph{Nature Reviews Materials}
  \textbf{2017}, \emph{2}, 1--15\relax
\mciteBstWouldAddEndPuncttrue
\mciteSetBstMidEndSepPunct{\mcitedefaultmidpunct}
{\mcitedefaultendpunct}{\mcitedefaultseppunct}\relax
\EndOfBibitem
\bibitem[Sokolikova and Mattevi(2020)Sokolikova, and
  Mattevi]{sokolikova2020direct}
Sokolikova,~M.~S.; Mattevi,~C. Direct synthesis of metastable phases of 2D
  transition metal dichalcogenides. \emph{Chemical Society Reviews}
  \textbf{2020}, \emph{49}, 3952--3980\relax
\mciteBstWouldAddEndPuncttrue
\mciteSetBstMidEndSepPunct{\mcitedefaultmidpunct}
{\mcitedefaultendpunct}{\mcitedefaultseppunct}\relax
\EndOfBibitem
\bibitem[Wang \latin{et~al.}(2018)Wang, Yu, Zhou, Li, Wong, Luo, Gan, and
  Zhai]{wang2018strategies}
Wang,~R.; Yu,~Y.; Zhou,~S.; Li,~H.; Wong,~H.; Luo,~Z.; Gan,~L.; Zhai,~T.
  Strategies on phase control in transition metal dichalcogenides.
  \emph{Advanced Functional Materials} \textbf{2018}, \emph{28}, 1802473\relax
\mciteBstWouldAddEndPuncttrue
\mciteSetBstMidEndSepPunct{\mcitedefaultmidpunct}
{\mcitedefaultendpunct}{\mcitedefaultseppunct}\relax
\EndOfBibitem
\bibitem[Yu \latin{et~al.}(2018)Yu, Nam, He, Wu, Zhang, Yang, Chen, Ma, Zhao,
  Liu, \latin{et~al.} others]{yu2018high}
Yu,~Y.; Nam,~G.-H.; He,~Q.; Wu,~X.-J.; Zhang,~K.; Yang,~Z.; Chen,~J.; Ma,~Q.;
  Zhao,~M.; Liu,~Z., \latin{et~al.}  High phase-purity 1T'-MoS$_2$ and
  1T'-MoSe$_2$ layered crystals. \emph{Nature Chemistry} \textbf{2018},
  \emph{10}, 638--643\relax
\mciteBstWouldAddEndPuncttrue
\mciteSetBstMidEndSepPunct{\mcitedefaultmidpunct}
{\mcitedefaultendpunct}{\mcitedefaultseppunct}\relax
\EndOfBibitem
\bibitem[Sandoval \latin{et~al.}(1991)Sandoval, Yang, Frindt, and
  Irwin]{sandoval1991raman}
Sandoval,~S.~J.; Yang,~D.; Frindt,~R.; Irwin,~J. Raman study and lattice
  dynamics of single molecular layers of MoS$_2$. \emph{Physical Review B}
  \textbf{1991}, \emph{44}, 3955\relax
\mciteBstWouldAddEndPuncttrue
\mciteSetBstMidEndSepPunct{\mcitedefaultmidpunct}
{\mcitedefaultendpunct}{\mcitedefaultseppunct}\relax
\EndOfBibitem
\bibitem[Huang \latin{et~al.}(2018)Huang, Fan, Singh, and
  Zheng]{huang2018first}
Huang,~H.; Fan,~X.; Singh,~D.~J.; Zheng,~W. First principles study on 2H-1T'
  transition in MoS$_2$ with copper. \emph{Physical Chemistry Chemical Physics}
  \textbf{2018}, \emph{20}, 26986--26994\relax
\mciteBstWouldAddEndPuncttrue
\mciteSetBstMidEndSepPunct{\mcitedefaultmidpunct}
{\mcitedefaultendpunct}{\mcitedefaultseppunct}\relax
\EndOfBibitem
\bibitem[Lin \latin{et~al.}(2014)Lin, Dumcenco, Huang, and
  Suenaga]{lin2014atomic}
Lin,~Y.-C.; Dumcenco,~D.~O.; Huang,~Y.-S.; Suenaga,~K. Atomic mechanism of the
  semiconducting-to-metallic phase transition in single-layered MoS$_2$.
  \emph{Nature Nanotechnology} \textbf{2014}, \emph{9}, 391--396\relax
\mciteBstWouldAddEndPuncttrue
\mciteSetBstMidEndSepPunct{\mcitedefaultmidpunct}
{\mcitedefaultendpunct}{\mcitedefaultseppunct}\relax
\EndOfBibitem
\bibitem[Hou \latin{et~al.}(2022)Hou, Zhang, Peng, Zhou, Wu, Xie, and
  Fang]{hou2022phase}
Hou,~X.; Zhang,~W.; Peng,~J.; Zhou,~L.; Wu,~J.; Xie,~K.; Fang,~Z. Phase
  Transformation of 1T'-MoS$_2$ Induced by Electrochemical Prelithiation for
  Lithium-Ion Storage. \emph{ACS Applied Energy Materials} \textbf{2022},
  \emph{5}, 11292--11303\relax
\mciteBstWouldAddEndPuncttrue
\mciteSetBstMidEndSepPunct{\mcitedefaultmidpunct}
{\mcitedefaultendpunct}{\mcitedefaultseppunct}\relax
\EndOfBibitem
\bibitem[Xu \latin{et~al.}(2020)Xu, Dai, Gaines, Hu, Tukker, and
  Steubing]{xu2020future}
Xu,~C.; Dai,~Q.; Gaines,~L.; Hu,~M.; Tukker,~A.; Steubing,~B. Future material
  demand for automotive lithium-based batteries. \emph{Communications
  Materials} \textbf{2020}, \emph{1}, 99\relax
\mciteBstWouldAddEndPuncttrue
\mciteSetBstMidEndSepPunct{\mcitedefaultmidpunct}
{\mcitedefaultendpunct}{\mcitedefaultseppunct}\relax
\EndOfBibitem
\bibitem[Hajiahmadi \latin{et~al.}(2023)Hajiahmadi, Ghasemi, Kühne, and
  Naghavi]{hajiahmadi2023first}
Hajiahmadi,~Z.; Ghasemi,~S.~A.; Kühne,~T.~D.; Naghavi,~S.~S. \emph{ACS Applied
  Nano Materials} \textbf{2023}, \emph{6}, 12862--12870\relax
\mciteBstWouldAddEndPuncttrue
\mciteSetBstMidEndSepPunct{\mcitedefaultmidpunct}
{\mcitedefaultendpunct}{\mcitedefaultseppunct}\relax
\EndOfBibitem
\bibitem[Sun \latin{et~al.}(2015)Sun, Wei, Li, Luo, An, Sheng, Wang, Chen, and
  Mai]{sun2015vanadium}
Sun,~R.; Wei,~Q.; Li,~Q.; Luo,~W.; An,~Q.; Sheng,~J.; Wang,~D.; Chen,~W.;
  Mai,~L. Vanadium sulfide on reduced graphene oxide layer as a promising anode
  for sodium ion battery. \emph{ACS Applied Materials \& Interfaces}
  \textbf{2015}, \emph{7}, 20902--20908\relax
\mciteBstWouldAddEndPuncttrue
\mciteSetBstMidEndSepPunct{\mcitedefaultmidpunct}
{\mcitedefaultendpunct}{\mcitedefaultseppunct}\relax
\EndOfBibitem
\bibitem[Dua \latin{et~al.}(2021)Dua, Deb, Paul, and Sarkar]{dua2021twin}
Dua,~H.; Deb,~J.; Paul,~D.; Sarkar,~U. Twin-graphene as a promising anode
  material for Na-ion rechargeable batteries. \emph{ACS Applied Nano Materials}
  \textbf{2021}, \emph{4}, 4912--4918\relax
\mciteBstWouldAddEndPuncttrue
\mciteSetBstMidEndSepPunct{\mcitedefaultmidpunct}
{\mcitedefaultendpunct}{\mcitedefaultseppunct}\relax
\EndOfBibitem
\bibitem[Deb \latin{et~al.}(2022)Deb, Ahuja, and Sarkar]{deb2022two}
Deb,~J.; Ahuja,~R.; Sarkar,~U. Two-dimensional pentagraphyne as a
  high-performance anode material for Li/Na-ion rechargeable batteries.
  \emph{ACS Applied Nano Materials} \textbf{2022}, \emph{5}, 10572--10582\relax
\mciteBstWouldAddEndPuncttrue
\mciteSetBstMidEndSepPunct{\mcitedefaultmidpunct}
{\mcitedefaultendpunct}{\mcitedefaultseppunct}\relax
\EndOfBibitem
\bibitem[Lin \latin{et~al.}(2019)Lin, Lei, Zhang, Liu, Wallace, and
  Chen]{lin2019two}
Lin,~L.; Lei,~W.; Zhang,~S.; Liu,~Y.; Wallace,~G.~G.; Chen,~J. Two-dimensional
  transition metal dichalcogenides in supercapacitors and secondary batteries.
  \emph{Energy Storage Materials} \textbf{2019}, \emph{19}, 408--423\relax
\mciteBstWouldAddEndPuncttrue
\mciteSetBstMidEndSepPunct{\mcitedefaultmidpunct}
{\mcitedefaultendpunct}{\mcitedefaultseppunct}\relax
\EndOfBibitem
\bibitem[Upadhyay \latin{et~al.}(2021)Upadhyay, Satrughna, and
  Pakhira]{upadhyay2021recent}
Upadhyay,~S.~N.; Satrughna,~J. A.~K.; Pakhira,~S. Recent advancements of
  two-dimensional transition metal dichalcogenides and their applications in
  electrocatalysis and energy storage. \emph{Emergent Materials} \textbf{2021},
  \emph{4}, 951--970\relax
\mciteBstWouldAddEndPuncttrue
\mciteSetBstMidEndSepPunct{\mcitedefaultmidpunct}
{\mcitedefaultendpunct}{\mcitedefaultseppunct}\relax
\EndOfBibitem
\bibitem[Stephenson \latin{et~al.}(2014)Stephenson, Li, Olsen, and
  Mitlin]{stephenson2014lithium}
Stephenson,~T.; Li,~Z.; Olsen,~B.; Mitlin,~D. Lithium ion battery applications
  of molybdenum disulfide (MoS$_2$) nanocomposites. \emph{Energy \&
  Environmental Science} \textbf{2014}, \emph{7}, 209--231\relax
\mciteBstWouldAddEndPuncttrue
\mciteSetBstMidEndSepPunct{\mcitedefaultmidpunct}
{\mcitedefaultendpunct}{\mcitedefaultseppunct}\relax
\EndOfBibitem
\bibitem[Barik and Pal(2019)Barik, and Pal]{barik2019defect}
Barik,~G.; Pal,~S. Defect induced performance enhancement of monolayer MoS$_2$
  for Li-and Na-ion batteries. \emph{The Journal of Physical Chemistry C}
  \textbf{2019}, \emph{123}, 21852--21865\relax
\mciteBstWouldAddEndPuncttrue
\mciteSetBstMidEndSepPunct{\mcitedefaultmidpunct}
{\mcitedefaultendpunct}{\mcitedefaultseppunct}\relax
\EndOfBibitem
\bibitem[He \latin{et~al.}(2022)He, Wang, Yin, Zhang, Chen, and
  Huang]{he20221t}
He,~X.; Wang,~R.; Yin,~H.; Zhang,~Y.; Chen,~W.; Huang,~S. 1T-MoS$_2$ monolayer
  as a promising anode material for (Li/Na/Mg)-ion batteries. \emph{Applied
  Surface Science} \textbf{2022}, \emph{584}, 152537\relax
\mciteBstWouldAddEndPuncttrue
\mciteSetBstMidEndSepPunct{\mcitedefaultmidpunct}
{\mcitedefaultendpunct}{\mcitedefaultseppunct}\relax
\EndOfBibitem
\bibitem[Hafner and Kresse(1997)Hafner, and Kresse]{VASP0}
Hafner,~J.; Kresse,~G. \emph{Properties of Complex Inorganic Solids}; Springer,
  1997; pp 69--82\relax
\mciteBstWouldAddEndPuncttrue
\mciteSetBstMidEndSepPunct{\mcitedefaultmidpunct}
{\mcitedefaultendpunct}{\mcitedefaultseppunct}\relax
\EndOfBibitem
\bibitem[Kresse and Furthm\"uller(1996)Kresse, and Furthm\"uller]{VASP1}
Kresse,~G.; Furthm\"uller,~J. Efficient iterative schemes for ab initio
  total-energy calculations using a plane-wave basis set. \emph{Physical Review
  B} \textbf{1996}, \emph{54}, 11169--11186\relax
\mciteBstWouldAddEndPuncttrue
\mciteSetBstMidEndSepPunct{\mcitedefaultmidpunct}
{\mcitedefaultendpunct}{\mcitedefaultseppunct}\relax
\EndOfBibitem
\bibitem[Kresse and Joubert(1999)Kresse, and Joubert]{VASP2}
Kresse,~G.; Joubert,~D. From ultrasoft pseudopotentials to the projector
  augmented-wave method. \emph{Physical Review B} \textbf{1999}, \emph{59},
  1758--1775\relax
\mciteBstWouldAddEndPuncttrue
\mciteSetBstMidEndSepPunct{\mcitedefaultmidpunct}
{\mcitedefaultendpunct}{\mcitedefaultseppunct}\relax
\EndOfBibitem
\bibitem[Perdew \latin{et~al.}(1996)Perdew, Burke, and Ernzerhof]{PBE}
Perdew,~J.~P.; Burke,~K.; Ernzerhof,~M. Generalized gradient approximation made
  simple. \emph{Physical Review Letters} \textbf{1996}, \emph{77}, 3865\relax
\mciteBstWouldAddEndPuncttrue
\mciteSetBstMidEndSepPunct{\mcitedefaultmidpunct}
{\mcitedefaultendpunct}{\mcitedefaultseppunct}\relax
\EndOfBibitem
\bibitem[Grimme \latin{et~al.}(2010)Grimme, Antony, Ehrlich, and
  Krieg]{grimme2010consistent}
Grimme,~S.; Antony,~J.; Ehrlich,~S.; Krieg,~H. A consistent and accurate ab
  initio parametrization of density functional dispersion correction (DFT-D)
  for the 94 elements H-Pu. \emph{The Journal of Chemical Physics}
  \textbf{2010}, \emph{132}, 154104\relax
\mciteBstWouldAddEndPuncttrue
\mciteSetBstMidEndSepPunct{\mcitedefaultmidpunct}
{\mcitedefaultendpunct}{\mcitedefaultseppunct}\relax
\EndOfBibitem
\bibitem[Grimme \latin{et~al.}(2011)Grimme, Ehrlich, and
  Goerigk]{grimme2011effect}
Grimme,~S.; Ehrlich,~S.; Goerigk,~L. Effect of the damping function in
  dispersion corrected density functional theory. \emph{Journal of
  Computational Chemistry} \textbf{2011}, \emph{32}, 1456--1465\relax
\mciteBstWouldAddEndPuncttrue
\mciteSetBstMidEndSepPunct{\mcitedefaultmidpunct}
{\mcitedefaultendpunct}{\mcitedefaultseppunct}\relax
\EndOfBibitem
\bibitem[Tang \latin{et~al.}(2009)Tang, Sanville, and Henkelman]{tang2009grid}
Tang,~W.; Sanville,~E.; Henkelman,~G. A grid-based Bader analysis algorithm
  without lattice bias. \emph{Journal of Physics: Condensed Matter}
  \textbf{2009}, \emph{21}, 084204\relax
\mciteBstWouldAddEndPuncttrue
\mciteSetBstMidEndSepPunct{\mcitedefaultmidpunct}
{\mcitedefaultendpunct}{\mcitedefaultseppunct}\relax
\EndOfBibitem
\bibitem[Sanville \latin{et~al.}(2007)Sanville, Kenny, Smith, and
  Henkelman]{sanville2007improved}
Sanville,~E.; Kenny,~S.~D.; Smith,~R.; Henkelman,~G. Improved grid-based
  algorithm for Bader charge allocation. \emph{Journal of computational
  chemistry} \textbf{2007}, \emph{28}, 899--908\relax
\mciteBstWouldAddEndPuncttrue
\mciteSetBstMidEndSepPunct{\mcitedefaultmidpunct}
{\mcitedefaultendpunct}{\mcitedefaultseppunct}\relax
\EndOfBibitem
\bibitem[Yu and Trinkle(2011)Yu, and Trinkle]{yu2011accurate}
Yu,~M.; Trinkle,~D.~R. Accurate and efficient algorithm for Bader charge
  integration. \emph{The Journal of Chemical Physics} \textbf{2011},
  \emph{134}, 064111\relax
\mciteBstWouldAddEndPuncttrue
\mciteSetBstMidEndSepPunct{\mcitedefaultmidpunct}
{\mcitedefaultendpunct}{\mcitedefaultseppunct}\relax
\EndOfBibitem
\bibitem[De~Souza \latin{et~al.}(2021)De~Souza, de~Castro, Marques, and
  Belchior]{de2021dft}
De~Souza,~L.; de~Castro,~G.~M.; Marques,~L.; Belchior,~J. A DFT investigation
  of lithium adsorption on graphenes as a potential anode material in
  lithium-ion batteries. \emph{Journal of Molecular Graphics and Modelling}
  \textbf{2021}, \emph{108}, 107998\relax
\mciteBstWouldAddEndPuncttrue
\mciteSetBstMidEndSepPunct{\mcitedefaultmidpunct}
{\mcitedefaultendpunct}{\mcitedefaultseppunct}\relax
\EndOfBibitem
\bibitem[Pajtler \latin{et~al.}(2023)Pajtler, Luka{\v{c}}evi{\'c},
  Du{\v{s}}i{\'c}, and Mu{\v{z}}evi{\'c}]{pajtler2023lithium}
Pajtler,~M.~V.; Luka{\v{c}}evi{\'c},~I.; Du{\v{s}}i{\'c},~V.;
  Mu{\v{z}}evi{\'c},~M. Lithium adsorption on the interface of graphene/boron
  nitride nanoribbons. \emph{Journal of Materials Science} \textbf{2023},
  \emph{58}, 4513--4524\relax
\mciteBstWouldAddEndPuncttrue
\mciteSetBstMidEndSepPunct{\mcitedefaultmidpunct}
{\mcitedefaultendpunct}{\mcitedefaultseppunct}\relax
\EndOfBibitem
\bibitem[Akgen{\c{c}}(2019)]{akgencc2019two}
Akgen{\c{c}},~B. Two-dimensional black arsenic for {Li}-ion battery
  applications: a {DFT} study. \emph{Journal of Materials Science}
  \textbf{2019}, \emph{54}, 9543--9552\relax
\mciteBstWouldAddEndPuncttrue
\mciteSetBstMidEndSepPunct{\mcitedefaultmidpunct}
{\mcitedefaultendpunct}{\mcitedefaultseppunct}\relax
\EndOfBibitem
\bibitem[Li \latin{et~al.}(2020)Li, Guo, and Jiao]{Ya_Meng_Li_2020}
Li,~Y.-M.; Guo,~Y.-L.; Jiao,~Z.-Y. The effect of {S}-functionalized and
  vacancies on {V$_2$C} {MXenes} as anode materials for {Na}-ion and {Li}-ion
  batteries. \emph{Current Applied Physics} \textbf{2020}, \emph{20},
  310--319\relax
\mciteBstWouldAddEndPuncttrue
\mciteSetBstMidEndSepPunct{\mcitedefaultmidpunct}
{\mcitedefaultendpunct}{\mcitedefaultseppunct}\relax
\EndOfBibitem
\bibitem[Zhou \latin{et~al.}(2004)Zhou, Cococcioni, Marianetti, Morgan, and
  Ceder]{zhou2004first}
Zhou,~F.; Cococcioni,~M.; Marianetti,~C.~A.; Morgan,~D.; Ceder,~G.
  First-principles prediction of redox potentials in transition-metal compounds
  with {LDA+U}. \emph{Physical Review B} \textbf{2004}, \emph{70}, 235121\relax
\mciteBstWouldAddEndPuncttrue
\mciteSetBstMidEndSepPunct{\mcitedefaultmidpunct}
{\mcitedefaultendpunct}{\mcitedefaultseppunct}\relax
\EndOfBibitem
\bibitem[Kim \latin{et~al.}(2020)Kim, Choi, Shin, Choi, Kim, Park, and
  Yoon]{kim2020applications}
Kim,~T.; Choi,~W.; Shin,~H.-C.; Choi,~J.-Y.; Kim,~J.~M.; Park,~M.-S.;
  Yoon,~W.-S. Applications of voltammetry in lithium ion battery research.
  \emph{Journal of Electrochemical Science and Technology} \textbf{2020},
  \emph{11}, 14--25\relax
\mciteBstWouldAddEndPuncttrue
\mciteSetBstMidEndSepPunct{\mcitedefaultmidpunct}
{\mcitedefaultendpunct}{\mcitedefaultseppunct}\relax
\EndOfBibitem
\bibitem[Gonz{\'a}lez \latin{et~al.}(2022)Gonz{\'a}lez, Vizcaya, and
  Morell]{gonzalez2022v}
Gonz{\'a}lez,~J.~W.; Vizcaya,~S.; Morell,~E.~S. $V_2C$-based lithium batteries:
  The influence of magnetic phase and Hubbard interaction. \emph{Functional
  Materials Letters} \textbf{2022}, 2340023\relax
\mciteBstWouldAddEndPuncttrue
\mciteSetBstMidEndSepPunct{\mcitedefaultmidpunct}
{\mcitedefaultendpunct}{\mcitedefaultseppunct}\relax
\EndOfBibitem
\bibitem[Soler \latin{et~al.}(2002)Soler, Artacho, Gale, Garc{\'\i}a, Junquera,
  Ordej{\'o}n, and S{\'a}nchez-Portal]{soler2002siesta}
Soler,~J.~M.; Artacho,~E.; Gale,~J.~D.; Garc{\'\i}a,~A.; Junquera,~J.;
  Ordej{\'o}n,~P.; S{\'a}nchez-Portal,~D. The SIESTA method for ab initio
  order-N materials simulation. \emph{Journal of Physics: Condensed Matter}
  \textbf{2002}, \emph{14}, 2745\relax
\mciteBstWouldAddEndPuncttrue
\mciteSetBstMidEndSepPunct{\mcitedefaultmidpunct}
{\mcitedefaultendpunct}{\mcitedefaultseppunct}\relax
\EndOfBibitem
\bibitem[Klime{\v{s}} \latin{et~al.}(2009)Klime{\v{s}}, Bowler, and
  Michaelides]{klimevs2009chemical}
Klime{\v{s}},~J.; Bowler,~D.~R.; Michaelides,~A. Chemical accuracy for the van
  der Waals density functional. \emph{Journal of Physics: Condensed Matter}
  \textbf{2009}, \emph{22}, 022201\relax
\mciteBstWouldAddEndPuncttrue
\mciteSetBstMidEndSepPunct{\mcitedefaultmidpunct}
{\mcitedefaultendpunct}{\mcitedefaultseppunct}\relax
\EndOfBibitem
\bibitem[Chen \latin{et~al.}(2021)Chen, Deng, Yan, Shi, Chang, Ding, Sun, Yang,
  and Liu]{chen2021diverse}
Chen,~K.; Deng,~J.; Yan,~Y.; Shi,~Q.; Chang,~T.; Ding,~X.; Sun,~J.; Yang,~S.;
  Liu,~J.~Z. Diverse electronic and magnetic properties of CrS$_2$ enabling
  strain-controlled 2D lateral heterostructure spintronic devices. \emph{NPJ
  Computational Materials} \textbf{2021}, \emph{7}, 79\relax
\mciteBstWouldAddEndPuncttrue
\mciteSetBstMidEndSepPunct{\mcitedefaultmidpunct}
{\mcitedefaultendpunct}{\mcitedefaultseppunct}\relax
\EndOfBibitem
\bibitem[Zhao \latin{et~al.}(2018)Zhao, Pan, Fang, Che, Wang, Bu, and
  Huang]{zhao2018metastable}
Zhao,~W.; Pan,~J.; Fang,~Y.; Che,~X.; Wang,~D.; Bu,~K.; Huang,~F. Metastable
  MoS$_2$: crystal structure, electronic band structure, synthetic approach and
  intriguing physical properties. \emph{Chemistry--A European Journal}
  \textbf{2018}, \emph{24}, 15942--15954\relax
\mciteBstWouldAddEndPuncttrue
\mciteSetBstMidEndSepPunct{\mcitedefaultmidpunct}
{\mcitedefaultendpunct}{\mcitedefaultseppunct}\relax
\EndOfBibitem
\bibitem[Xu \latin{et~al.}(2018)Xu, Yan, and Qiao]{xu2018magnetism}
Xu,~W.; Yan,~S.; Qiao,~W. Magnetism in monolayer 1T-MoS$_2$ and 1T-MoS$_2$H
  tuned by strain. \emph{RSC Advances} \textbf{2018}, \emph{8},
  8435--8441\relax
\mciteBstWouldAddEndPuncttrue
\mciteSetBstMidEndSepPunct{\mcitedefaultmidpunct}
{\mcitedefaultendpunct}{\mcitedefaultseppunct}\relax
\EndOfBibitem
\bibitem[Urban \latin{et~al.}(2016)Urban, Seo, and
  Ceder]{urban2016computational}
Urban,~A.; Seo,~D.-H.; Ceder,~G. Computational understanding of Li-ion
  batteries. \emph{npj Computational Materials} \textbf{2016}, \emph{2},
  1--13\relax
\mciteBstWouldAddEndPuncttrue
\mciteSetBstMidEndSepPunct{\mcitedefaultmidpunct}
{\mcitedefaultendpunct}{\mcitedefaultseppunct}\relax
\EndOfBibitem
\bibitem[Liu \latin{et~al.}(2021)Liu, Zhang, Chen, Huang, Zhang, Zhou, Du, and
  Xiao]{liu2021modulating}
Liu,~K.; Zhang,~B.; Chen,~X.; Huang,~Y.; Zhang,~P.; Zhou,~D.; Du,~H.; Xiao,~B.
  Modulating the open-circuit voltage of two-dimensional {MoB} {MBene}
  electrode via specific surface chemistry for {Na/K} ion batteries: A
  first-principles study. \emph{The Journal of Physical Chemistry C}
  \textbf{2021}, \emph{125}, 18098--18107\relax
\mciteBstWouldAddEndPuncttrue
\mciteSetBstMidEndSepPunct{\mcitedefaultmidpunct}
{\mcitedefaultendpunct}{\mcitedefaultseppunct}\relax
\EndOfBibitem
\bibitem[Persson \latin{et~al.}(2010)Persson, Hinuma, Meng, Van~der Ven, and
  Ceder]{persson2010thermodynamic}
Persson,~K.; Hinuma,~Y.; Meng,~Y.~S.; Van~der Ven,~A.; Ceder,~G. Thermodynamic
  and kinetic properties of the Li-graphite system from first-principles
  calculations. \emph{Physical Review B} \textbf{2010}, \emph{82}, 125416\relax
\mciteBstWouldAddEndPuncttrue
\mciteSetBstMidEndSepPunct{\mcitedefaultmidpunct}
{\mcitedefaultendpunct}{\mcitedefaultseppunct}\relax
\EndOfBibitem
\bibitem[Sukhanova \latin{et~al.}(2022)Sukhanova, Bereznikova, Manakhov,
  Al~Qahtani, and Popov]{sukhanova2022novel}
Sukhanova,~E.~V.; Bereznikova,~L.~A.; Manakhov,~A.~M.; Al~Qahtani,~H.~S.;
  Popov,~Z.~I. A novel membrane-like 2D A'-MoS$_2$ as anode for Lithium-and
  Sodium-ion batteries. \emph{Membranes} \textbf{2022}, \emph{12}, 1156\relax
\mciteBstWouldAddEndPuncttrue
\mciteSetBstMidEndSepPunct{\mcitedefaultmidpunct}
{\mcitedefaultendpunct}{\mcitedefaultseppunct}\relax
\EndOfBibitem
\bibitem[Artacho \latin{et~al.}(2008)Artacho, Anglada, Di{\'e}guez, Gale,
  Garc{\'\i}a, Junquera, Martin, Ordej{\'o}n, Pruneda, S{\'a}nchez-Portal, and
  Soler]{artacho2008siesta}
Artacho,~E.; Anglada,~E.; Di{\'e}guez,~O.; Gale,~J.~D.; Garc{\'\i}a,~A.;
  Junquera,~J.; Martin,~R.~M.; Ordej{\'o}n,~P.; Pruneda,~J.~M.;
  S{\'a}nchez-Portal,~D.; Soler,~J.~M. The SIESTA method; developments and
  applicability. \emph{Journal of Physics: Condensed Matter} \textbf{2008},
  \emph{20}, 064208\relax
\mciteBstWouldAddEndPuncttrue
\mciteSetBstMidEndSepPunct{\mcitedefaultmidpunct}
{\mcitedefaultendpunct}{\mcitedefaultseppunct}\relax
\EndOfBibitem
\bibitem[Xiong \latin{et~al.}(2011)Xiong, Slater, Balasubramanian, Johnson, and
  Rajh]{xiong2011amorphous}
Xiong,~H.; Slater,~M.~D.; Balasubramanian,~M.; Johnson,~C.~S.; Rajh,~T.
  Amorphous TiO$_2$ nanotube anode for rechargeable sodium ion batteries.
  \emph{The Journal of Physical Chemistry Letters} \textbf{2011}, \emph{2},
  2560--2565\relax
\mciteBstWouldAddEndPuncttrue
\mciteSetBstMidEndSepPunct{\mcitedefaultmidpunct}
{\mcitedefaultendpunct}{\mcitedefaultseppunct}\relax
\EndOfBibitem
\bibitem[Qin \latin{et~al.}(2015)Qin, Chen, Pan, Niu, Hu, Li, Li, and
  Sun]{qin2015mos2}
Qin,~W.; Chen,~T.; Pan,~L.; Niu,~L.; Hu,~B.; Li,~D.; Li,~J.; Sun,~Z.
  MoS$_2$-reduced graphene oxide composites via microwave assisted synthesis
  for sodium ion battery anode with improved capacity and cycling performance.
  \emph{Electrochimica Acta} \textbf{2015}, \emph{153}, 55--61\relax
\mciteBstWouldAddEndPuncttrue
\mciteSetBstMidEndSepPunct{\mcitedefaultmidpunct}
{\mcitedefaultendpunct}{\mcitedefaultseppunct}\relax
\EndOfBibitem
\bibitem[Kumar~Gudelli and Guo(2019)Kumar~Gudelli, and Guo]{kumar2019magnetism}
Kumar~Gudelli,~V.; Guo,~G.-Y. Magnetism and magneto-optical effects in bulk and
  few-layer CrI$_3$: a theoretical GGA+ U study. \emph{New Journal of Physics}
  \textbf{2019}, \emph{21}, 053012\relax
\mciteBstWouldAddEndPuncttrue
\mciteSetBstMidEndSepPunct{\mcitedefaultmidpunct}
{\mcitedefaultendpunct}{\mcitedefaultseppunct}\relax
\EndOfBibitem
\end{mcitethebibliography}

\providecommand{\latin}[1]{#1}
\makeatletter
\providecommand{\doi}
  {\begingroup\let\do\@makeother\dospecials
  \catcode`\{=1 \catcode`\}=2 \doi@aux}
\providecommand{\doi@aux}[1]{\endgroup\texttt{#1}}
\makeatother
\providecommand*\mcitethebibliography{\thebibliography}
\csname @ifundefined\endcsname{endmcitethebibliography}
  {\let\endmcitethebibliography\endthebibliography}{}

\end{document}